\documentclass[%
aps,prl,twocolumn,superscriptaddress,groupedaddress
 amsmath,amssymb,
footinbib,
 reprint,%
]{revtex4-2}

\usepackage{graphicx}
\usepackage{dcolumn}
\usepackage{bm}
\usepackage{textgreek}
\usepackage[utf8]{inputenc}
\usepackage[T1]{fontenc}
\usepackage{mathptmx}
\usepackage{upgreek}
\usepackage{hyperref}

\usepackage{color}

\usepackage{natbib}
\usepackage{amsmath}

\newcommand{\beginsupplement}{%
        \setcounter{table}{0}
        \renewcommand{\thetable}{S\arabic{table}}%
        \setcounter{figure}{0}
        \renewcommand{\thefigure}{S\arabic{figure}}%
				\renewcommand{\theequation}{S.\arabic{equation}}
     }

\begin{document}

\title[CNT and NbSe2]{Supercurrent and phase slips in a ballistic carbon nanotube embedded into a van der Waals heterostructure} 

\author{Christian B\"auml}
\author{Lorenz Bauriedl}
\affiliation{Institut f\"ur Experimentelle und Angewandte Physik, University of Regensburg}
\author{Magdalena Marganska}
\author{Milena Grifoni}
\affiliation{Institut f\"ur Theoretische Physik, University of Regensburg}
\author{Christoph Strunk}
\author{Nicola Paradiso}\email{nicola.paradiso@physik.uni-regensburg.de}
\affiliation{Institut f\"ur Experimentelle und Angewandte Physik, University of Regensburg}



\begin{abstract}	
We demonstrate long-range superconducting correlations in a several micrometer-long carbon nanotube encapsulated in a van der Waals stack between hBN and NbSe$_2$. We show that a substantial supercurrent flows through the nanotube section beneath the NbSe$_2$ crystal as well as  through the 2~\textmu m-long section not in contact with it.  As expected for superconductors of nanoscopic cross section, the current-induced breakdown of superconductivity is characterized by resistance steps due to the nucleation of phase slip centers. All elements of our hybrid device are active building blocks of several recently proposed setups for realization of Majorana fermions in carbon  nanotubes.
\end{abstract}

\maketitle

Among the many schemes proposed~\cite{Lutchyn2010,Oreg2010} for the demonstration of Majorana fermions (MFs), the most popular consists of a 1D superconductor with large spin-orbit interaction (SOI) in the presence of a magnetic field. It was in such systems that first signatures of Majorana quasiparticles were reported~\cite{Mourik1003,Deng2012,Das2012,Churchill2013,Finck2013,Deng1557,Lutchyn2018}. The 1D conductor is typically a III-V semiconducting nanowire proximitized by Al. In the last years, several works~\cite{Egger2012,Sau2013,Hsu2015,Marganska2018,Milz2019,Lesser2020} have pointed out the advantage to use carbon nanotubes (CNTs) instead. They are, in fact, genuinely one dimensional~\cite{Saitobook}, since their cross section comprises only a handful of atoms: as a consequence, they have well-separated transverse modes.  Also, clean CNTs  provide ballistic transport~\cite{Laird2014} and, owing to their curvature, show a sizable spin-orbit field parallel to their axis~\cite{Izumida2009,Kuemmeth2008,Churchill2009,Jhang2011,Jespersen2011,Steele2013}.  From the theoretical point of view, their spectral properties can be suitably computed at a microscopic level.

So far, clean transport features in CNTs were mainly demonstrated on suspended carbon nanotubes~\cite{Cao2005,Wu2010,Pei2012,Ranjan2015,Waissman2013,Gramich2015,Blien2018}. However, recent theoretical works~\cite{Marganska2018,Lesser2020} have proposed the combination of CNTs with superconducting transition metal dichalcogenides (such as, e.g.,~NbSe$_2$). This is clearly incompatible with the disorder-free scheme based on suspended CNTs. This situation is somehow similar to that of graphene one decade ago: high mobility devices required graphene suspended above the substrate~\cite{Meyer2007,Du2008}, while the observation of new physics often required the combination of graphene with other materials. The impasse was resolved by the introduction of hBN encapsulation~\cite{Wang614}, a crucial qualitative leap in the fabrication of the so-called van der Waals heterostructures~\cite{Geim2013,Novoselovaac9439}.

In this work, we apply ideas and methods of van der Waals stacking to CNTs, demonstrating a hybrid 1D-2D superconducting heterostructure.  We investigate a CNT encapsulated between  a few-layer NbSe$_2$ crystal and a hBN flake on a graphite substrate. We demonstrate that superconducting correlations extends throughout the whole CNT length, including the few micrometer-long portions next to the NbSe$_2$ crystal. 
At finite current bias, a series of resistance steps is observed, indicating the nucleation of phase slip centers, as expected for a quantum wire of nanometric cross section.

\begin{figure*}[tb]
\includegraphics[width=2\columnwidth]{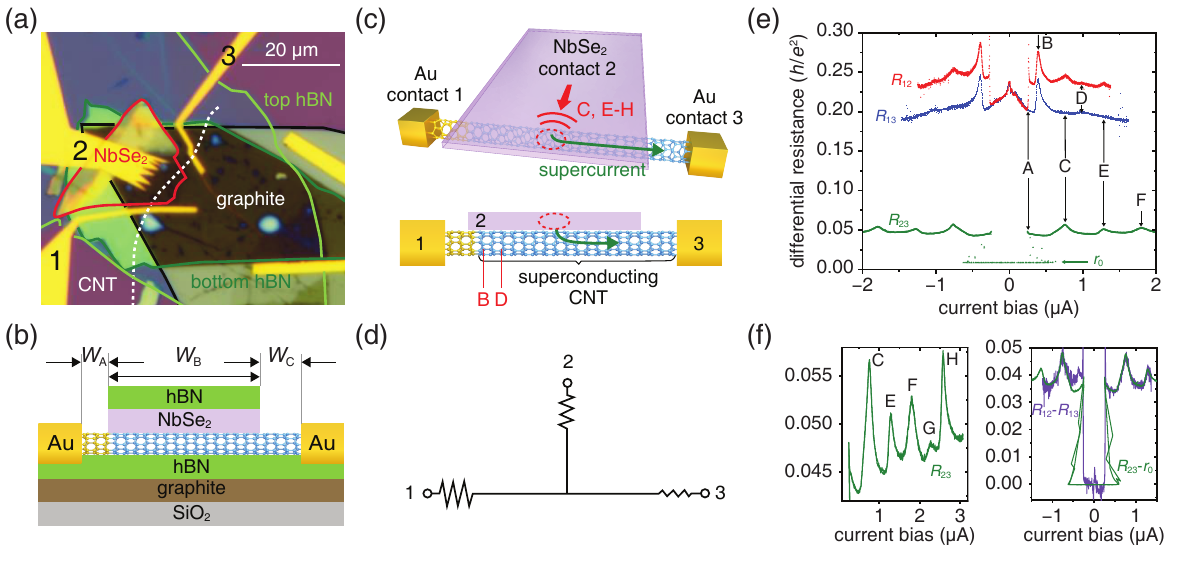}
\caption{(a) Optical micrograph of the final device. The boundaries of the different crystals are indicated by the color lines. Electrodes 1 and 3 contact the carbon nanotube (CNT) while 2 contacts the few-layer NbSe$_2$ crystal. (b) Stacking scheme of the van der Waals heterostructure. The lengths of the three CNT portions are $W_A=0.2$~\textmu m, $W_B=7.5$~\textmu m, and $W_C=2$~\textmu m. (c) Schematics representing the location of phase slip lines and centers. A region of higher transparency (ellipse) between CNT and NbSe$_2$ implements an analogue of the point contact experiment in Ref.~\cite{Paradiso_2DMat_2019}. The phase slip centers (see text) producing the resistance peaks B and D are located somewhere in the CNT portion between NbSe$_2$ and contact 1. The region between NbSe$_2$ and contact 3 appears to be perfectly superconducting, with no signature of resistance. (d) Electrical replacement scheme of the device. (e) Two-terminal differential resistance versus current bias between contacts 1-2 ($R_{12}$, red), 1-3 ($R_{13}$, blue) and 2--3 ($R_{23}$, green). A zoom-in of the latter graph is shown in (f). (g) Plot of the difference between $R_{12}$ and $R_{13}$, together with $R_{23}-r_0$, where $r_0$ is the residual low-bias resistance in $R_{23}$. The perfect match supports the electrical scheme shown in (d).
}
\label{fig:firstfig}
\end{figure*}



Our device is fabricated starting from a flake of graphite and one of hBN exfoliated and stamped on a standard Si/SiO$_2$ chip using the dry stamping technique described in Ref.~\cite{Castellanos_Gomez_2014}. As in Ref.~\cite{Overweg2018PRL,Overweg2018NL} the graphite/hBN  stack provides an atomically flat conductive backgate capable to screen charge traps at the SiO$_2$ surface as well as disorder from contaminants near the CNT. This is due to the close proximity between CNT and graphite, which is given by the thickness ($\approx 5$~nm) of the bottom hBN layer.

On a separate chip, macroscopically long CNTs are grown using a standard recipe for clean single wall CNTs described elsewhere~\cite{Blien2018}. The growth chip with CNTs is then imaged by scanning electron microscopy and a suitable CNT is picked up by a polydimethylsiloxane (PDMS) stamp coated with a polycarbonate (PC) layer, following the standard technique~\cite{Wang614} used in van der Waals heterostructure fabrication.
The application of the pick-up method to CNTs has been reported by Huang \textit{et al.}~\cite{HuangCNTedge}. Unlike these authors, we observe a weak adhesion between CNT and hBN, therefore we pick up the CNT directly by the PC/PDMS stamp. After dissolving the PC in chloroform, a topography scan by atomic force microscopy (AFM) revealed no significant trace of contamination. Then, pick-up stamping in N$_2$ atmosphere is used to deposit a few-layer NbSe$_2$ flake (capped with a thicker hBN crystal) onto the CNT. Gold  edge contacts to CNT and NbSe$_2$ are fabricated simultaneously using the recipe established for encapsulated graphene~\cite{Wang614}. Despite the CNT-Au interface being limited to a handful of atoms, very transparent contacts can be obtained which support currents of hundreds of microamperes. Further information about sample fabrication is provided in the Supplementary Material~\footnote{See Supplementary Material for further information}.

The completed sample is shown in Fig.~\ref{fig:firstfig}(a)  and illustrated schematically in Fig.~\ref{fig:firstfig}(b,c).
It has  three terminals: two are Au edge contacts and one is provided by the NbSe$_2$ flake in contact with the CNT along a distance of $W_B=7.5$~\textmu m. 
This central part is located in between two uncovered CNT portions whose length is $W_A=0.2$~\textmu m and $W_C=2$~\textmu m, respectively [see Fig.~\ref{fig:firstfig}(b)]. 
The sample is mounted on the cold finger of a dilution refrigerator with carefully filtered DC transport lines. 
We measure two-terminal differential conductance at finite bias for different temperatures or magnetic fields. Current is measured between the pairs of contacts 1--2, 2--3 or 1--3, see Fig.~\ref{fig:firstfig}(c). The comparison between the conductance measurement in the three configurations disentangles contributions from different parts of the device.
Figure~\ref{fig:firstfig}(e) shows the current dependence of the differential resistance $R_{ij}= \partial V_{ij}/\partial I_{ij}$, where $V_{ij}$ and $I_{ij}$ are the voltage and the current between the contacts $i$ and $j$, respectively. The measurement is performed at base temperature ($T=30$~mK) and in absence of magnetic field. From Fig.~\ref{fig:firstfig}(e), we deduce that, in good approximation, $R_{12}\approx R_{13}+ R_{23}$, see also panel (h). More precisely $R_{12}-R_{13}=R_{23}-r_0$ where $r_0$ is the low-bias value of $R_{23}$. From the electrical point of view, this corresponds to a three-terminal circuit sketched in Fig.~\ref{fig:firstfig}(d), where the resistance associated to the contact 3 is negligible.

All the traces show steps in resistance which are separated by peaks, as shown in Fig.~\ref{fig:firstfig}(e). The position of these  features is indeed controlled by current and not by voltage bias. In fact, they are nicely aligned only when plotted as a function of current~\cite{Note1}. Moreover, they occur for voltage bias larger than any relevant energy scale, e.g.~the gap of bulk NbSe$_2$, which is approximately 1~meV. This excludes spectroscopic features such as Andreev reflection or tunneling as possible source of the observed conductance modulation.
The measured resistance traces appear very similar to those observed in the literature~\cite{Kasumov1508,KasumovPRB2003,Ferrier2004} for short CNT-based Josephson junctions and are attributed to phase slip events, as discussed below.
At high bias, the resistance is of the order of 0.2 $h/e^2$ between contacts 1--2 and 1--3, while it is much lower between contacts 2--3, i.e.~about $0.05$~$h/e^2$. This corresponds to a conductance of $\approx 20$~$e^2/h\approx$~(1.3k\textOmega)$^{-1}$, i.e., five times the theoretical limit for a single undoped CNT (that is,  $4e^2/h$ per spin- and valley-degenerate channel). A reason for such a large conductance could be that we picked up indeed a bundle of few CNTs. This large conductance is nevertheless small compared to the impressive value at low bias: for current below 600~nA the two terminal resistance is $R_{23}\equiv r_0=0.0087$~$h/e^2$, or 225~\textOmega, which corresponds to a conductance of nearly $120$~$e^2/h$. 
Indeed the device conductance is much larger than that, since, out of 225~\textOmega, at least 80~\textOmega~are due to the cryostat cables while the Au-NbSe$_2$ contact resistance is unknown, but estimated to be of the order of 100~\textOmega~based on our previous experience with similar contacts. Therefore, we estimate the two-terminal conductance of the device  to be at most of the order of tens of ohms, corresponding to a conductance several hundreds of $e^2/h$. In the normal state, this conductance would require a many-nm-thick bundle of CNTs. By AFM scans~\cite{Note1} we measured a CNT diameter of approximately 1~nm, excluding a thick bundle of many CNTs. The presence of only a few CNTs is in agreement with the  high temperature state conductance of the CNT which is of the order of 18~$e^2/h$, see Fig.~\ref{fig:firstfig}(e). Finally, this large conductance is current-independent up to the critical value~\cite{Note1}. This suggests to attribute the small residual $r_0$ entirely to cryostat cables and Au contact resistance.

The observations above indicate that the large low-bias conductance is produced by a finite supercurrent between contacts 2--3. The supercurrent and the phase slip resistance steps unambiguously show that the CNT is superconducting.    Since any normal-conducting CNT section between the contacts 2 and 3 would limit the two-terminal conductance to $4e^2/h$ per spin- and valley-degenerate channel, we  conclude that the CNT must be superconducting  up to the very interface with the Au edge contact 3.

If the CNT is superconducting, we expect to observe the peculiar characteristics of 1D superconductors, namely, current-driven steps in conductance due to nucleation of phase slip centers~\cite{ARUTYUNOV20081,IvlevKopninAdv1984,Bezryadin2013,Michotte2004,Vodolazov2007}. Such steps were reported since the very first observation of supercurrent in CNT-based Josephson junctions~\cite{Kasumov1508,KasumovPRB2003}. On the other hand, at sufficiently high current densities, transport characteristics in few-layer NbSe$_2$ show similar step-like traces due to nucleation of phase slip \textit{lines} (PSL)~\cite{Paradiso_2DMat_2019,Tran,SaitoPRMat,Moriya2020,Li2020}. To discriminate between PSCs and PSLs, we need to carefully compare the three traces in Fig.~\ref{fig:firstfig}(e). As shown below, we can  approximately locate the position of the sources of the dissipative features. 
From the data we can conclude that: ($i$) the features B, D indicate phase slip centers \textit{within} the CNT, located in the 1--2 portion, as they are visible in the 1--2 and 1--3 traces but not in 2--3. ($ii$) The features C, E, F, G and H are located on the NbSe$_2$ side, (visible in the 2--3 and 1--2 traces but absent in 1--3). We interpret these steps as PSLs in NbSe$_2$.  The relatively low current necessary to trigger the PSL nucleation (compared to Ref.~\cite{Paradiso_2DMat_2019}) suggests that there is a relatively narrow portion of CNT where the contact to NbSe$_2$ is particularly transparent (see sketch in Fig.~\ref{fig:firstfig}(c)).
($iii$) The interpretation of the feature A, namely the supercurrent jump,  is more difficult, but of crucial importance. For current bias above 600~nA the resistance $R_{23}$ acquires a finite value of nearly 0.05 $h/e^2$, which is, as discussed above, approximately the difference between $R_{12}$ and $R_{23}$. This value is basically the contact resistance between the multimode NbSe$_2$ and the few-mode CNT, which is the only source of resistance in a perfectly ballistic quantum wire. Notice that   \textit{all} measured resistive features in $R_{23}$ are also present in the difference between $R_{12}$ and $R_{13}$ (see Fig.~\ref{fig:firstfig}(g)), indicating that no relevant scattering processes are taking place in the portion 2-3. 

The supercurrent between contacts 2 and 3 is flowing through the 2~\textmu m-long CNT segment $W_C$ not in contact with NbSe$_2$. To explain a supercurrent in this segment there are two possibilities: either the superconductivity in this region is proximity-induced or intrinsic. The former case is the most intuitive: the transparent NbSe$_2$-CNT interface induces by proximity effect superconducting correlations in the nearby CNT segment. This \textit{lateral} proximity effect can extend up to a length $L_N=\min (L_T, L_{\phi})$, where $L_T= \hbar v_F/(k_BT)$ is the thermal length in the ballistic case, and $L_{\phi}$ is the coherence length in the CNT. Assuming for the Fermi velocity $v_F=10^6$~m/s, we find that for $T<1$~K, $L_T>7.7$~\textmu m,   compatible with our device size. 
The absence of normal conducting regions in the CNT portion between 2 and 3 implies that the Au contact only reduces, by inverse proximity,  the order parameter on the CNT side of the Au-CNT interface, without fully suppressing it. 

\begin{figure*}[tb]
\includegraphics[width=2\columnwidth]{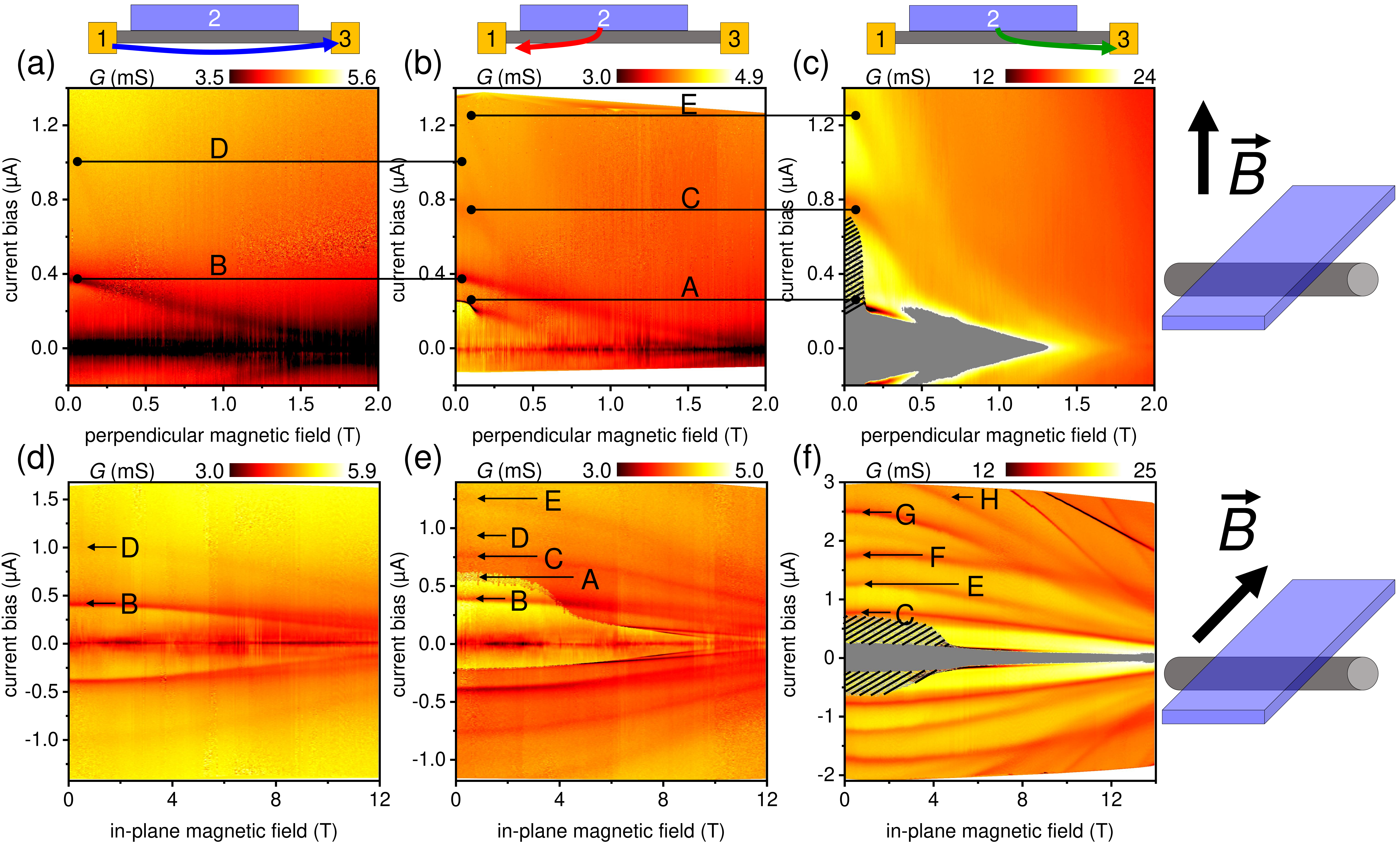}
\caption{The graphs show the differential conductance between contacts 1--3 (a,d), 1--2 (b,e) and 2--3 (c,f) as a function of current bias and perpendicular (a,b,c) or in-plane (d,e,f) magnetic field~\cite{Note2}. The in-plane magnetic field is approximately perpendicular to the CNT axis. In the first row the graphs have the same scale for the current axis. We observe that the  features have the same current and field dependence in the different contact configurations. This allows us to identify the same feature in different graphs and assign to it the same label (horizontal lines). In (c,f) conductance saturating beyond  25~mS is depicted in grey. Dashed areas in (c,f) are regions corresponding to the situation where to one current bias value corresponds to two conductance values. This is due to the fact that we have a voltage resistance connected to a voltage source, see text.} 
\label{fig:allcolor}
\end{figure*}

The alternative scenario is that of an intrinsic superconductivity of the CNT, which was reported~\cite{KasumovPRB2003,Kociak2001,Ferrier2004,Tang2462,Takesue2006,Shi2012} for thick \textit{bundles} of CNTs. Indeed, the bundle nature of the CNTs is suggested by the large high-bias (normal) conductance between contacts 2-3, which indicates the presence of several channels, as discussed above. Sufficiently long CNT bundles have been reported to be intrinsically superconducting~\cite{Ferrier2004,Shi2012}, with a large variety of critical temperatures up to several kelvins. However, all the reported bundles contained a large number of CNTs, of the order of many tens up to hundreds, clearly incompatible with our case. Besides, the intrinsic superconductivity scenario can be excluded by comparing conductance measurements in-plane and out-of-plane magnetic field, as discussed below.

\begin{figure}[t!]
\includegraphics[width=\columnwidth]{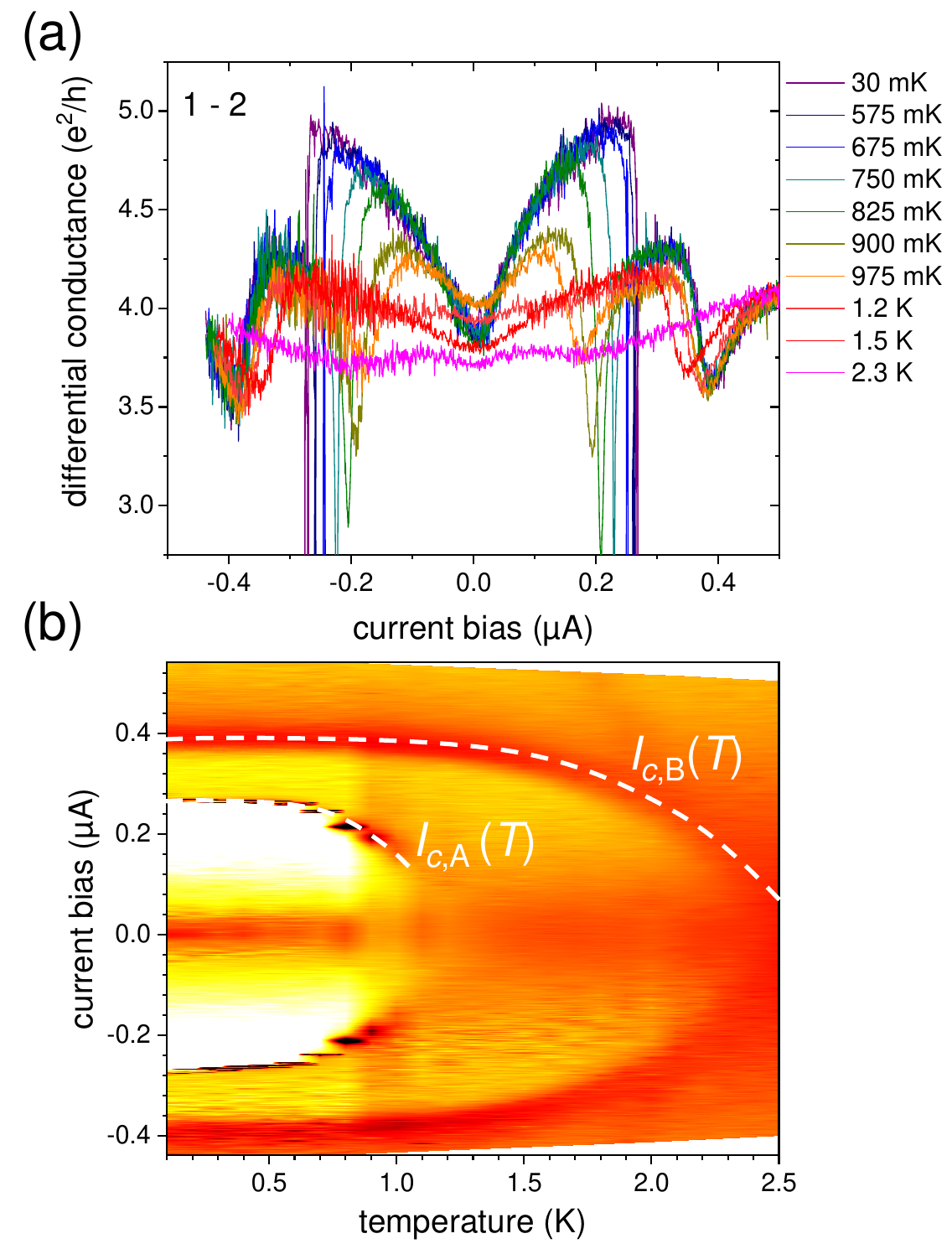}
\caption{(a) Differential conductance as a function of current bias between contacts 1--2, plotted for selected temperatures. (b) Color plot of the differential conductance between contacts 1--2, plotted as a function of current and temperature. The plot clearly shows  the temperature dependence of the critical current bias  $I_{\text{A}}$ and $I_{\text{B}}$ triggering  the feature A and B, respectively.
}
\label{fig:Tdep}
\end{figure}


Based on the comparison of the resistance traces between contacts 1--2, 1--3 and 2--3 in Fig.~\ref{fig:firstfig}(e), we argued above that the features (B,D) and (C,E,F,G,H) originate from  PSCs and PSLs,  respectively. On the other hand, resistance steps produced by PSCs in CNTs and PSLs in NbSe$_2$ are expected to differ in their perpendicular magnetic field dependence. In particular, the current triggering a PSC in a CNT decreases almost linearly with the magnetic field, and the associated resistance peak is visible up to fields comparable to the critical field in NbSe$_2$~\cite{Kasumov1508,KasumovPRB2003}. In contrast, the current triggering a PSL weakly depends on the magnetic field for low fields, displaying a downward curvature, and already at moderate fields the PSL feature is washed-out altogether~\cite{Paradiso_2DMat_2019}.


To corroborate the above attribution, we have performed finite-bias conductance measurements as a function of in- and out-of-plane magnetic field.  Panels  (a), (b) and (c) in Fig.~\ref{fig:allcolor} refer respectively to contact configuration 1--3, 1--2, and 2--3, measured with magnetic field applied \textit{out-of-plane}, while panels  (d), (e) and (f) refer to the same respective contact configuration,  measured with magnetic field applied \textit{in-plane}~\footnote{The conductance measurements with in-plane magnetic field (Fig.~\ref{fig:allcolor}(d,e,f)) were performed in a second cool-down, necessary to rotate the sample.}. 


As shown in Fig.~\ref{fig:allcolor}(a)  feature B is visible up to 2~T, where it merges with the zero-bias conductance dip (which is discussed below). It decreases almost linearly with the magnetic field, similarly to the conductance steps in short CNT Josephson junctions~\cite{Kasumov1508,KasumovPRB2003,Ferrier2004}.  The feature D is instead very faint and cannot be properly discerned. 

Interestingly, the features  C, E, F, G, and H, which we attributed to PSLs, disappear at $B_{\perp}\approx 0.3$~T (see Fig.~\ref{fig:allcolor}(b,c)). This is consistent with the behavior observed~\cite{Paradiso_2DMat_2019} in plain few-layer NbSe$_2$ and predicted by time-dependent Ginzburg-Landau simulations by Berdiyorov \textit{et al.}~\cite{Berdiyorov2009,Berdiyorov2014}, who attributed the smearing to the introduction of Abrikosov vortices.

On the other hand, application of an in-plane field \textit{does not} wash out such PSL-related features, since no significant amount of vortices is introduced  by an in-plane field as the NbSe$_2$ thickness is smaller that the coherence length.  Instead, the in-plane field simply reduces the gap and correspondingly the current thresholds for the nucleation of PSLs. Since NbSe$_2$ is an Ising superconductor (singlet Cooper pairs protected by spin-valley locking), one needs very large parallel fields to quench the superconducting gap, much larger than the maximum field available in our setup~\cite{Xi2016,delaBarrera2018,Xing2017}.  As a consequence, in Fig.~\ref{fig:allcolor}(f)  the PSL conductance dips are observed for lower bias at high field, but the reduction is modest even for the largest magnetic field experimentally accessible, i.e.~14~T.

The behavior of the feature B in out-of-plane and in-plane magnetic field [shown in Figs.~\ref{fig:allcolor}(a,d)] rules out the possibility of intrinsic superconductivity in the CNT. In fact, from the CNT point of view in both cases the magnetic field vector is perpendicular to the axis, therefore one would expect no difference in the magnetic field dependence. It is evident that the striking critical field enhancement for this feature in Fig.~\ref{fig:allcolor}(d) reflects the Ising  behavior of the parent superconductor.

We stress that in our measurements we apply a voltage bias: when applied between contacts 1--2 or 1--3, the resistance of contact 1 is sufficiently large that current and voltage bias are roughly proportional. However, for contacts 2--3 the device resistance is so low, that at the critical value for the supercurrent the resistance increase produces a marked decrease of current. The heating power produces a jump to the overheated part of the voltage-current characteristics. Thus, the $R_{23}(I)$ curve is not single-valued, as shown in Fig.~\ref{fig:firstfig}(e,g). In the color plots of Fig.~\ref{fig:allcolor}(c,f) such two-valued regions (dashed in the graphs) cannot be properly represented.

A comment is in order about the relatively broad zero bias resistance peak (conductance dip) between terminal 1--2 and 1--3. This peak seems to be related to a potential barrier of finite height and width at the contact 1. Such zero-bias anomaly has been observed also in short CNT Josephson junctions~\cite{KasumovPRB2003}; it is not related to superconductivity: it persists, in fact, up to magnetic fields well beyond the critical value, as shown above.

Finally, we turn to the temperature dependence of the features discussed above.  Figure~\ref{fig:Tdep}(a) shows, for selected temperatures, the conductance versus bias for the contact configuration 1--2. The measured current range allows us to follow the temperature dependence of the features A and B. Clearly, by increasing temperature their critical currents $I_{c,\text{A}}$ and $I_{c,\text{B}}$ decrease and their visibility fades out. However, the temperature dependence is different for the two features. This is better visible in the corresponding  color plot of Fig.~\ref{fig:Tdep}(b), where the zero-bias dip, the features A and B are clearly visible. Interestingly, the critical current $I_{c,\text{B}}(T)$ has a temperature dependence very similar to that reported for suspended proximitized CNTs~\cite{Kasumov1508,KasumovPRB2003}, i.e.,~a weak dependence at low temperature and a pronounced decrease close to a critical temperature, which is $\simeq 2.6$~K in our case. We emphasize the remarkable similarity of both magnetic field (linear dependence as shown by Fig.~\ref{fig:allcolor}(a)) and temperature dependence of $I_{c,\text{B}}$ with the corresponding feature observed in short suspended CNTs, see e.g., Fig.~4B in Ref.~\cite{Kasumov1508} and Fig.~9 in Ref.~\cite{KasumovPRB2003}.

The critical value for the supercurrent (feature A) shows an approximately similar temperature dependence, but with a $T_c$ that can be extrapolated towards approximately 1.5~K, i.e., about half of that for feature B, see Fig.~\ref{fig:Tdep}(b). We shall thus consider $\Delta^{\ast}=1.764k_B\cdot 1.5$~K=230~\textmu eV as the gap for the superconducting CNT. 
If $I_{c,\text{A}}$ represents the critical current of the Josephson junction 
between the NbSe$_2$ crystal (with gap $\Delta_{\text{NbSe2}}=390$~\textmu eV) and the proximitized CNT (with gap $\Delta^{\ast}$), then the low-temperature limit of the  Ambegaokar-Baratoff formula~\cite{AB,ABerratum}  $I_c R_N=\Delta^{\ast}K(\sqrt{1-(\Delta^{\ast}/\Delta_{\text{NbSe2}})^2})$ must hold (here $K(x)$ represents the elliptic integral of the first kind~\cite{AB,ABerratum}).
 The measured product $I_cR_N$ is about 0.77~mV ($I_c=I_{c,\text{A}}(0)=600$~nA) while $ \Delta^{\ast}K(\sqrt{1-(\Delta^{\ast}/\Delta_{\text{NbSe2}})^2}) \approx$ 0.7~mV, using $e\Delta^{\ast}=1.764k_B\times \mbox{1.5~K}$ and  $e\Delta_{\text{NbSe}}=1.764k_B\times \mbox{5.5~K}$. Within the uncertainty in both $I_{c,\text{A}}$ and $\Delta^{\ast}$, the two values match.

In conclusion, we have demonstrated a long-range lateral proximity effect in a hybrid 1D-2D van der Waals heterostructure consisting of NbSe$_2$ and a 9 \textmu m-long carbon nanotube. The low level of disorder and the high contact transparency enable the observation of a large supercurrent through the carbon nanotube, including the 2 \textmu m-long section not directly in contact with NbSe$_2$. Finite current bias triggers distinctive resistance peaks, which we identify as phase slip centers,  clear signatures of superconductivity within the carbon nanotube.


\begin{acknowledgments}
The work was funded by the Deutsche Forschungsgemeinschaft (DFG, German Research Foundation) – Project-ID 
314695032 – SFB 1277 (Subprojects B04 and B08), and by the European Union’s Horizon 2020 research and innovation programme
under grant agreements No 862660 QUANTUM E-LEAPS.
\end{acknowledgments}

\bibliography{biblio1}

\begin{thebibliography}{65}%
\makeatletter
\providecommand \@ifxundefined [1]{%
 \@ifx{#1\undefined}
}%
\providecommand \@ifnum [1]{%
 \ifnum #1\expandafter \@firstoftwo
 \else \expandafter \@secondoftwo
 \fi
}%
\providecommand \@ifx [1]{%
 \ifx #1\expandafter \@firstoftwo
 \else \expandafter \@secondoftwo
 \fi
}%
\providecommand \natexlab [1]{#1}%
\providecommand \enquote  [1]{``#1''}%
\providecommand \bibnamefont  [1]{#1}%
\providecommand \bibfnamefont [1]{#1}%
\providecommand \citenamefont [1]{#1}%
\providecommand \href@noop [0]{\@secondoftwo}%
\providecommand \href [0]{\begingroup \@sanitize@url \@href}%
\providecommand \@href[1]{\@@startlink{#1}\@@href}%
\providecommand \@@href[1]{\endgroup#1\@@endlink}%
\providecommand \@sanitize@url [0]{\catcode `\\12\catcode `\$12\catcode
  `\&12\catcode `\#12\catcode `\^12\catcode `\_12\catcode `\%12\relax}%
\providecommand \@@startlink[1]{}%
\providecommand \@@endlink[0]{}%
\providecommand \url  [0]{\begingroup\@sanitize@url \@url }%
\providecommand \@url [1]{\endgroup\@href {#1}{\urlprefix }}%
\providecommand \urlprefix  [0]{URL }%
\providecommand \Eprint [0]{\href }%
\providecommand \doibase [0]{https://doi.org/}%
\providecommand \selectlanguage [0]{\@gobble}%
\providecommand \bibinfo  [0]{\@secondoftwo}%
\providecommand \bibfield  [0]{\@secondoftwo}%
\providecommand \translation [1]{[#1]}%
\providecommand \BibitemOpen [0]{}%
\providecommand \bibitemStop [0]{}%
\providecommand \bibitemNoStop [0]{.\EOS\space}%
\providecommand \EOS [0]{\spacefactor3000\relax}%
\providecommand \BibitemShut  [1]{\csname bibitem#1\endcsname}%
\let\auto@bib@innerbib\@empty
\bibitem [{\citenamefont {Lutchyn}\ \emph {et~al.}(2010)\citenamefont
  {Lutchyn}, \citenamefont {Sau},\ and\ \citenamefont
  {Das~Sarma}}]{Lutchyn2010}%
  \BibitemOpen
  \bibfield  {author} {\bibinfo {author} {\bibfnamefont {R.~M.}\ \bibnamefont
  {Lutchyn}}, \bibinfo {author} {\bibfnamefont {J.~D.}\ \bibnamefont {Sau}},\
  and\ \bibinfo {author} {\bibfnamefont {S.}~\bibnamefont {Das~Sarma}},\
  }\bibfield  {title} {\bibinfo {title} {{Majorana Fermions and a Topological
  Phase Transition in Semiconductor-Superconductor Heterostructures}},\ }\href
  {https://doi.org/10.1103/PhysRevLett.105.077001} {\bibfield  {journal}
  {\bibinfo  {journal} {Phys. Rev. Lett.}\ }\textbf {\bibinfo {volume} {105}},\
  \bibinfo {pages} {077001} (\bibinfo {year} {2010})}\BibitemShut {NoStop}%
\bibitem [{\citenamefont {Oreg}\ \emph {et~al.}(2010)\citenamefont {Oreg},
  \citenamefont {Refael},\ and\ \citenamefont {von Oppen}}]{Oreg2010}%
  \BibitemOpen
  \bibfield  {author} {\bibinfo {author} {\bibfnamefont {Y.}~\bibnamefont
  {Oreg}}, \bibinfo {author} {\bibfnamefont {G.}~\bibnamefont {Refael}},\ and\
  \bibinfo {author} {\bibfnamefont {F.}~\bibnamefont {von Oppen}},\ }\bibfield
  {title} {\bibinfo {title} {{Helical Liquids and Majorana Bound States in
  Quantum Wires}},\ }\href {https://doi.org/10.1103/PhysRevLett.105.177002}
  {\bibfield  {journal} {\bibinfo  {journal} {Phys. Rev. Lett.}\ }\textbf
  {\bibinfo {volume} {105}},\ \bibinfo {pages} {177002} (\bibinfo {year}
  {2010})}\BibitemShut {NoStop}%
\bibitem [{\citenamefont {Mourik}\ \emph {et~al.}(2012)\citenamefont {Mourik},
  \citenamefont {Zuo}, \citenamefont {Frolov}, \citenamefont {Plissard},
  \citenamefont {Bakkers},\ and\ \citenamefont {Kouwenhoven}}]{Mourik1003}%
  \BibitemOpen
  \bibfield  {author} {\bibinfo {author} {\bibfnamefont {V.}~\bibnamefont
  {Mourik}}, \bibinfo {author} {\bibfnamefont {K.}~\bibnamefont {Zuo}},
  \bibinfo {author} {\bibfnamefont {S.~M.}\ \bibnamefont {Frolov}}, \bibinfo
  {author} {\bibfnamefont {S.~R.}\ \bibnamefont {Plissard}}, \bibinfo {author}
  {\bibfnamefont {E.~P. A.~M.}\ \bibnamefont {Bakkers}},\ and\ \bibinfo
  {author} {\bibfnamefont {L.~P.}\ \bibnamefont {Kouwenhoven}},\ }\bibfield
  {title} {\bibinfo {title} {{Signatures of Majorana Fermions in Hybrid
  Superconductor-Semiconductor Nanowire Devices}},\ }\href
  {https://doi.org/10.1126/science.1222360} {\bibfield  {journal} {\bibinfo
  {journal} {Science}\ }\textbf {\bibinfo {volume} {336}},\ \bibinfo {pages}
  {1003} (\bibinfo {year} {2012})}\BibitemShut {NoStop}%
\bibitem [{\citenamefont {Deng}\ \emph {et~al.}(2012)\citenamefont {Deng},
  \citenamefont {Yu}, \citenamefont {Huang}, \citenamefont {Larsson},
  \citenamefont {Caroff},\ and\ \citenamefont {Xu}}]{Deng2012}%
  \BibitemOpen
  \bibfield  {author} {\bibinfo {author} {\bibfnamefont {M.~T.}\ \bibnamefont
  {Deng}}, \bibinfo {author} {\bibfnamefont {C.~L.}\ \bibnamefont {Yu}},
  \bibinfo {author} {\bibfnamefont {G.~Y.}\ \bibnamefont {Huang}}, \bibinfo
  {author} {\bibfnamefont {M.}~\bibnamefont {Larsson}}, \bibinfo {author}
  {\bibfnamefont {P.}~\bibnamefont {Caroff}},\ and\ \bibinfo {author}
  {\bibfnamefont {H.~Q.}\ \bibnamefont {Xu}},\ }\bibfield  {title} {\bibinfo
  {title} {{Anomalous Zero-Bias Conductance Peak in a Nb-InSb Nanowire-Nb
  Hybrid Device}},\ }\href {https://doi.org/10.1021/nl303758w} {\bibfield
  {journal} {\bibinfo  {journal} {Nano Letters}\ }\textbf {\bibinfo {volume}
  {12}},\ \bibinfo {pages} {6414} (\bibinfo {year} {2012})}\BibitemShut
  {NoStop}%
\bibitem [{\citenamefont {Das}\ \emph {et~al.}(2012)\citenamefont {Das},
  \citenamefont {Ronen}, \citenamefont {Most}, \citenamefont {Oreg},
  \citenamefont {Heiblum},\ and\ \citenamefont {Shtrikman}}]{Das2012}%
  \BibitemOpen
  \bibfield  {author} {\bibinfo {author} {\bibfnamefont {A.}~\bibnamefont
  {Das}}, \bibinfo {author} {\bibfnamefont {Y.}~\bibnamefont {Ronen}}, \bibinfo
  {author} {\bibfnamefont {Y.}~\bibnamefont {Most}}, \bibinfo {author}
  {\bibfnamefont {Y.}~\bibnamefont {Oreg}}, \bibinfo {author} {\bibfnamefont
  {M.}~\bibnamefont {Heiblum}},\ and\ \bibinfo {author} {\bibfnamefont
  {H.}~\bibnamefont {Shtrikman}},\ }\bibfield  {title} {\bibinfo {title}
  {{Zero-bias peaks and splitting in an Al-InAs nanowire topological
  superconductor as a signature of Majorana fermions}},\ }\href
  {https://doi.org/10.1038/nphys2479} {\bibfield  {journal} {\bibinfo
  {journal} {Nature Physics}\ }\textbf {\bibinfo {volume} {8}},\ \bibinfo
  {pages} {887} (\bibinfo {year} {2012})}\BibitemShut {NoStop}%
\bibitem [{\citenamefont {Churchill}\ \emph {et~al.}(2013)\citenamefont
  {Churchill}, \citenamefont {Fatemi}, \citenamefont {Grove-Rasmussen},
  \citenamefont {Deng}, \citenamefont {Caroff}, \citenamefont {Xu},\ and\
  \citenamefont {Marcus}}]{Churchill2013}%
  \BibitemOpen
  \bibfield  {author} {\bibinfo {author} {\bibfnamefont {H.~O.~H.}\
  \bibnamefont {Churchill}}, \bibinfo {author} {\bibfnamefont {V.}~\bibnamefont
  {Fatemi}}, \bibinfo {author} {\bibfnamefont {K.}~\bibnamefont
  {Grove-Rasmussen}}, \bibinfo {author} {\bibfnamefont {M.~T.}\ \bibnamefont
  {Deng}}, \bibinfo {author} {\bibfnamefont {P.}~\bibnamefont {Caroff}},
  \bibinfo {author} {\bibfnamefont {H.~Q.}\ \bibnamefont {Xu}},\ and\ \bibinfo
  {author} {\bibfnamefont {C.~M.}\ \bibnamefont {Marcus}},\ }\bibfield  {title}
  {\bibinfo {title} {{Superconductor-nanowire devices from tunneling to the
  multichannel regime: Zero-bias oscillations and magnetoconductance
  crossover}},\ }\href {https://doi.org/10.1103/PhysRevB.87.241401} {\bibfield
  {journal} {\bibinfo  {journal} {Phys. Rev. B}\ }\textbf {\bibinfo {volume}
  {87}},\ \bibinfo {pages} {241401} (\bibinfo {year} {2013})}\BibitemShut
  {NoStop}%
\bibitem [{\citenamefont {Finck}\ \emph {et~al.}(2013)\citenamefont {Finck},
  \citenamefont {Van~Harlingen}, \citenamefont {Mohseni}, \citenamefont
  {Jung},\ and\ \citenamefont {Li}}]{Finck2013}%
  \BibitemOpen
  \bibfield  {author} {\bibinfo {author} {\bibfnamefont {A.~D.~K.}\
  \bibnamefont {Finck}}, \bibinfo {author} {\bibfnamefont {D.~J.}\ \bibnamefont
  {Van~Harlingen}}, \bibinfo {author} {\bibfnamefont {P.~K.}\ \bibnamefont
  {Mohseni}}, \bibinfo {author} {\bibfnamefont {K.}~\bibnamefont {Jung}},\ and\
  \bibinfo {author} {\bibfnamefont {X.}~\bibnamefont {Li}},\ }\bibfield
  {title} {\bibinfo {title} {{Anomalous Modulation of a Zero-Bias Peak in a
  Hybrid Nanowire-Superconductor Device}},\ }\href
  {https://doi.org/10.1103/PhysRevLett.110.126406} {\bibfield  {journal}
  {\bibinfo  {journal} {Phys. Rev. Lett.}\ }\textbf {\bibinfo {volume} {110}},\
  \bibinfo {pages} {126406} (\bibinfo {year} {2013})}\BibitemShut {NoStop}%
\bibitem [{\citenamefont {Deng}\ \emph {et~al.}(2016)\citenamefont {Deng},
  \citenamefont {Vaitiekenas}, \citenamefont {Hansen}, \citenamefont {Danon},
  \citenamefont {Leijnse}, \citenamefont {Flensberg}, \citenamefont {Nyg{\r
  a}rd}, \citenamefont {Krogstrup},\ and\ \citenamefont {Marcus}}]{Deng1557}%
  \BibitemOpen
  \bibfield  {author} {\bibinfo {author} {\bibfnamefont {M.~T.}\ \bibnamefont
  {Deng}}, \bibinfo {author} {\bibfnamefont {S.}~\bibnamefont {Vaitiekenas}},
  \bibinfo {author} {\bibfnamefont {E.~B.}\ \bibnamefont {Hansen}}, \bibinfo
  {author} {\bibfnamefont {J.}~\bibnamefont {Danon}}, \bibinfo {author}
  {\bibfnamefont {M.}~\bibnamefont {Leijnse}}, \bibinfo {author} {\bibfnamefont
  {K.}~\bibnamefont {Flensberg}}, \bibinfo {author} {\bibfnamefont
  {J.}~\bibnamefont {Nyg{\r a}rd}}, \bibinfo {author} {\bibfnamefont
  {P.}~\bibnamefont {Krogstrup}},\ and\ \bibinfo {author} {\bibfnamefont
  {C.~M.}\ \bibnamefont {Marcus}},\ }\bibfield  {title} {\bibinfo {title}
  {Majorana bound state in a coupled quantum-dot hybrid-nanowire system},\
  }\href {https://doi.org/10.1126/science.aaf3961} {\bibfield  {journal}
  {\bibinfo  {journal} {Science}\ }\textbf {\bibinfo {volume} {354}},\ \bibinfo
  {pages} {1557} (\bibinfo {year} {2016})}\BibitemShut {NoStop}%
\bibitem [{\citenamefont {Lutchyn}\ \emph {et~al.}(2018)\citenamefont
  {Lutchyn}, \citenamefont {Bakkers}, \citenamefont {Kouwenhoven},
  \citenamefont {Krogstrup}, \citenamefont {Marcus},\ and\ \citenamefont
  {Oreg}}]{Lutchyn2018}%
  \BibitemOpen
  \bibfield  {author} {\bibinfo {author} {\bibfnamefont {R.~M.}\ \bibnamefont
  {Lutchyn}}, \bibinfo {author} {\bibfnamefont {E.~P. A.~M.}\ \bibnamefont
  {Bakkers}}, \bibinfo {author} {\bibfnamefont {L.~P.}\ \bibnamefont
  {Kouwenhoven}}, \bibinfo {author} {\bibfnamefont {P.}~\bibnamefont
  {Krogstrup}}, \bibinfo {author} {\bibfnamefont {C.~M.}\ \bibnamefont
  {Marcus}},\ and\ \bibinfo {author} {\bibfnamefont {Y.}~\bibnamefont {Oreg}},\
  }\bibfield  {title} {\bibinfo {title} {Majorana zero modes in
  superconductor--semiconductor heterostructures},\ }\href
  {https://doi.org/10.1038/s41578-018-0003-1} {\bibfield  {journal} {\bibinfo
  {journal} {Nature Reviews Materials}\ }\textbf {\bibinfo {volume} {3}},\
  \bibinfo {pages} {52} (\bibinfo {year} {2018})}\BibitemShut {NoStop}%
\bibitem [{\citenamefont {Egger}\ and\ \citenamefont
  {Flensberg}(2012)}]{Egger2012}%
  \BibitemOpen
  \bibfield  {author} {\bibinfo {author} {\bibfnamefont {R.}~\bibnamefont
  {Egger}}\ and\ \bibinfo {author} {\bibfnamefont {K.}~\bibnamefont
  {Flensberg}},\ }\bibfield  {title} {\bibinfo {title} {{Emerging Dirac and
  Majorana fermions for carbon nanotubes with proximity-induced pairing and
  spiral magnetic field}},\ }\href {https://doi.org/10.1103/PhysRevB.85.235462}
  {\bibfield  {journal} {\bibinfo  {journal} {Phys. Rev. B}\ }\textbf {\bibinfo
  {volume} {85}},\ \bibinfo {pages} {235462} (\bibinfo {year}
  {2012})}\BibitemShut {NoStop}%
\bibitem [{\citenamefont {Sau}\ and\ \citenamefont {Tewari}(2013)}]{Sau2013}%
  \BibitemOpen
  \bibfield  {author} {\bibinfo {author} {\bibfnamefont {J.~D.}\ \bibnamefont
  {Sau}}\ and\ \bibinfo {author} {\bibfnamefont {S.}~\bibnamefont {Tewari}},\
  }\bibfield  {title} {\bibinfo {title} {{Topological superconducting state and
  Majorana fermions in carbon nanotubes}},\ }\href
  {https://doi.org/10.1103/PhysRevB.88.054503} {\bibfield  {journal} {\bibinfo
  {journal} {Phys. Rev. B}\ }\textbf {\bibinfo {volume} {88}},\ \bibinfo
  {pages} {054503} (\bibinfo {year} {2013})}\BibitemShut {NoStop}%
\bibitem [{\citenamefont {Hsu}\ \emph {et~al.}(2015)\citenamefont {Hsu},
  \citenamefont {Stano}, \citenamefont {Klinovaja},\ and\ \citenamefont
  {Loss}}]{Hsu2015}%
  \BibitemOpen
  \bibfield  {author} {\bibinfo {author} {\bibfnamefont {C.-H.}\ \bibnamefont
  {Hsu}}, \bibinfo {author} {\bibfnamefont {P.}~\bibnamefont {Stano}}, \bibinfo
  {author} {\bibfnamefont {J.}~\bibnamefont {Klinovaja}},\ and\ \bibinfo
  {author} {\bibfnamefont {D.}~\bibnamefont {Loss}},\ }\bibfield  {title}
  {\bibinfo {title} {{Antiferromagnetic nuclear spin helix and topological
  superconductivity in $^{13}\text{C}$ nanotubes}},\ }\href
  {https://doi.org/10.1103/PhysRevB.92.235435} {\bibfield  {journal} {\bibinfo
  {journal} {Phys. Rev. B}\ }\textbf {\bibinfo {volume} {92}},\ \bibinfo
  {pages} {235435} (\bibinfo {year} {2015})}\BibitemShut {NoStop}%
\bibitem [{\citenamefont {Marganska}\ \emph {et~al.}(2018)\citenamefont
  {Marganska}, \citenamefont {Milz}, \citenamefont {Izumida}, \citenamefont
  {Strunk},\ and\ \citenamefont {Grifoni}}]{Marganska2018}%
  \BibitemOpen
  \bibfield  {author} {\bibinfo {author} {\bibfnamefont {M.}~\bibnamefont
  {Marganska}}, \bibinfo {author} {\bibfnamefont {L.}~\bibnamefont {Milz}},
  \bibinfo {author} {\bibfnamefont {W.}~\bibnamefont {Izumida}}, \bibinfo
  {author} {\bibfnamefont {C.}~\bibnamefont {Strunk}},\ and\ \bibinfo {author}
  {\bibfnamefont {M.}~\bibnamefont {Grifoni}},\ }\bibfield  {title} {\bibinfo
  {title} {Majorana quasiparticles in semiconducting carbon nanotubes},\ }\href
  {https://doi.org/10.1103/PhysRevB.97.075141} {\bibfield  {journal} {\bibinfo
  {journal} {Phys. Rev. B}\ }\textbf {\bibinfo {volume} {97}},\ \bibinfo
  {pages} {075141} (\bibinfo {year} {2018})}\BibitemShut {NoStop}%
\bibitem [{\citenamefont {Milz}\ \emph {et~al.}(2019)\citenamefont {Milz},
  \citenamefont {Izumida}, \citenamefont {Grifoni},\ and\ \citenamefont
  {Marganska}}]{Milz2019}%
  \BibitemOpen
  \bibfield  {author} {\bibinfo {author} {\bibfnamefont {L.}~\bibnamefont
  {Milz}}, \bibinfo {author} {\bibfnamefont {W.}~\bibnamefont {Izumida}},
  \bibinfo {author} {\bibfnamefont {M.}~\bibnamefont {Grifoni}},\ and\ \bibinfo
  {author} {\bibfnamefont {M.}~\bibnamefont {Marganska}},\ }\bibfield  {title}
  {\bibinfo {title} {{Transverse profile and three-dimensional spin canting of
  a Majorana state in carbon nanotubes}},\ }\href
  {https://doi.org/10.1103/PhysRevB.100.155417} {\bibfield  {journal} {\bibinfo
   {journal} {Phys. Rev. B}\ }\textbf {\bibinfo {volume} {100}},\ \bibinfo
  {pages} {155417} (\bibinfo {year} {2019})}\BibitemShut {NoStop}%
\bibitem [{\citenamefont {Lesser}\ \emph {et~al.}(2020)\citenamefont {Lesser},
  \citenamefont {Shavit},\ and\ \citenamefont {Oreg}}]{Lesser2020}%
  \BibitemOpen
  \bibfield  {author} {\bibinfo {author} {\bibfnamefont {O.}~\bibnamefont
  {Lesser}}, \bibinfo {author} {\bibfnamefont {G.}~\bibnamefont {Shavit}},\
  and\ \bibinfo {author} {\bibfnamefont {Y.}~\bibnamefont {Oreg}},\ }\bibfield
  {title} {\bibinfo {title} {Topological superconductivity in carbon nanotubes
  with a small magnetic flux},\ }\href
  {https://doi.org/10.1103/PhysRevResearch.2.023254} {\bibfield  {journal}
  {\bibinfo  {journal} {Phys. Rev. Research}\ }\textbf {\bibinfo {volume}
  {2}},\ \bibinfo {pages} {023254} (\bibinfo {year} {2020})}\BibitemShut
  {NoStop}%
\bibitem [{\citenamefont {Saito}\ \emph {et~al.}(1998)\citenamefont {Saito},
  \citenamefont {Dresselhaus},\ and\ \citenamefont {Dresselhaus}}]{Saitobook}%
  \BibitemOpen
  \bibfield  {author} {\bibinfo {author} {\bibfnamefont {R.}~\bibnamefont
  {Saito}}, \bibinfo {author} {\bibfnamefont {G.}~\bibnamefont {Dresselhaus}},\
  and\ \bibinfo {author} {\bibfnamefont {M.~S.}\ \bibnamefont {Dresselhaus}},\
  }\href {https://doi.org/10.1142/p080} {\emph {\bibinfo {title} {Physical
  Properties of Carbon Nanotubes}}}\ (\bibinfo  {publisher} {World
  Scientific},\ \bibinfo {year} {1998})\BibitemShut {NoStop}%
\bibitem [{\citenamefont {Laird}\ \emph {et~al.}(2015)\citenamefont {Laird},
  \citenamefont {Kuemmeth}, \citenamefont {Steele}, \citenamefont
  {Grove-Rasmussen}, \citenamefont {Nyg\aa{}rd}, \citenamefont {Flensberg},\
  and\ \citenamefont {Kouwenhoven}}]{Laird2014}%
  \BibitemOpen
  \bibfield  {author} {\bibinfo {author} {\bibfnamefont {E.~A.}\ \bibnamefont
  {Laird}}, \bibinfo {author} {\bibfnamefont {F.}~\bibnamefont {Kuemmeth}},
  \bibinfo {author} {\bibfnamefont {G.~A.}\ \bibnamefont {Steele}}, \bibinfo
  {author} {\bibfnamefont {K.}~\bibnamefont {Grove-Rasmussen}}, \bibinfo
  {author} {\bibfnamefont {J.}~\bibnamefont {Nyg\aa{}rd}}, \bibinfo {author}
  {\bibfnamefont {K.}~\bibnamefont {Flensberg}},\ and\ \bibinfo {author}
  {\bibfnamefont {L.~P.}\ \bibnamefont {Kouwenhoven}},\ }\bibfield  {title}
  {\bibinfo {title} {Quantum transport in carbon nanotubes},\ }\href
  {https://doi.org/10.1103/RevModPhys.87.703} {\bibfield  {journal} {\bibinfo
  {journal} {Rev. Mod. Phys.}\ }\textbf {\bibinfo {volume} {87}},\ \bibinfo
  {pages} {703} (\bibinfo {year} {2015})}\BibitemShut {NoStop}%
\bibitem [{\citenamefont {Izumida}\ \emph {et~al.}(2009)\citenamefont
  {Izumida}, \citenamefont {Sato},\ and\ \citenamefont {Saito}}]{Izumida2009}%
  \BibitemOpen
  \bibfield  {author} {\bibinfo {author} {\bibfnamefont {W.}~\bibnamefont
  {Izumida}}, \bibinfo {author} {\bibfnamefont {K.}~\bibnamefont {Sato}},\ and\
  \bibinfo {author} {\bibfnamefont {R.}~\bibnamefont {Saito}},\ }\bibfield
  {title} {\bibinfo {title} {{Spin–Orbit Interaction in Single Wall Carbon
  Nanotubes: Symmetry Adapted Tight-Binding Calculation and Effective Model
  Analysis}},\ }\href@noop {} {\bibfield  {journal} {\bibinfo  {journal}
  {Journal of the Physical Society of Japan}\ }\textbf {\bibinfo {volume}
  {78}},\ \bibinfo {pages} {074707} (\bibinfo {year} {2009})}\BibitemShut
  {NoStop}%
\bibitem [{\citenamefont {Kuemmeth}\ \emph {et~al.}(2008)\citenamefont
  {Kuemmeth}, \citenamefont {Ilani}, \citenamefont {Ralph},\ and\ \citenamefont
  {McEuen}}]{Kuemmeth2008}%
  \BibitemOpen
  \bibfield  {author} {\bibinfo {author} {\bibfnamefont {F.}~\bibnamefont
  {Kuemmeth}}, \bibinfo {author} {\bibfnamefont {S.}~\bibnamefont {Ilani}},
  \bibinfo {author} {\bibfnamefont {D.~C.}\ \bibnamefont {Ralph}},\ and\
  \bibinfo {author} {\bibfnamefont {P.~L.}\ \bibnamefont {McEuen}},\ }\bibfield
   {title} {\bibinfo {title} {Coupling of spin and orbital motion of electrons
  in carbon nanotubes},\ }\href {https://doi.org/10.1038/nature06822}
  {\bibfield  {journal} {\bibinfo  {journal} {Nature}\ }\textbf {\bibinfo
  {volume} {452}},\ \bibinfo {pages} {448} (\bibinfo {year}
  {2008})}\BibitemShut {NoStop}%
\bibitem [{\citenamefont {Churchill}\ \emph {et~al.}(2009)\citenamefont
  {Churchill}, \citenamefont {Kuemmeth}, \citenamefont {Harlow}, \citenamefont
  {Bestwick}, \citenamefont {Rashba}, \citenamefont {Flensberg}, \citenamefont
  {Stwertka}, \citenamefont {Taychatanapat}, \citenamefont {Watson},\ and\
  \citenamefont {Marcus}}]{Churchill2009}%
  \BibitemOpen
  \bibfield  {author} {\bibinfo {author} {\bibfnamefont {H.~O.~H.}\
  \bibnamefont {Churchill}}, \bibinfo {author} {\bibfnamefont {F.}~\bibnamefont
  {Kuemmeth}}, \bibinfo {author} {\bibfnamefont {J.~W.}\ \bibnamefont
  {Harlow}}, \bibinfo {author} {\bibfnamefont {A.~J.}\ \bibnamefont
  {Bestwick}}, \bibinfo {author} {\bibfnamefont {E.~I.}\ \bibnamefont
  {Rashba}}, \bibinfo {author} {\bibfnamefont {K.}~\bibnamefont {Flensberg}},
  \bibinfo {author} {\bibfnamefont {C.~H.}\ \bibnamefont {Stwertka}}, \bibinfo
  {author} {\bibfnamefont {T.}~\bibnamefont {Taychatanapat}}, \bibinfo {author}
  {\bibfnamefont {S.~K.}\ \bibnamefont {Watson}},\ and\ \bibinfo {author}
  {\bibfnamefont {C.~M.}\ \bibnamefont {Marcus}},\ }\bibfield  {title}
  {\bibinfo {title} {{Relaxation and Dephasing in a Two-Electron
  $^{13}\mathbf{C}$ Nanotube Double Quantum Dot}},\ }\href
  {https://doi.org/10.1103/PhysRevLett.102.166802} {\bibfield  {journal}
  {\bibinfo  {journal} {Phys. Rev. Lett.}\ }\textbf {\bibinfo {volume} {102}},\
  \bibinfo {pages} {166802} (\bibinfo {year} {2009})}\BibitemShut {NoStop}%
\bibitem [{\citenamefont {Jhang}\ \emph {et~al.}(2011)\citenamefont {Jhang},
  \citenamefont {Marga\ifmmode~\acute{n}\else \'{n}\fi{}ska}, \citenamefont
  {Skourski}, \citenamefont {Preusche}, \citenamefont {Grifoni}, \citenamefont
  {Wosnitza},\ and\ \citenamefont {Strunk}}]{Jhang2011}%
  \BibitemOpen
  \bibfield  {author} {\bibinfo {author} {\bibfnamefont {S.~H.}\ \bibnamefont
  {Jhang}}, \bibinfo {author} {\bibfnamefont {M.}~\bibnamefont
  {Marga\ifmmode~\acute{n}\else \'{n}\fi{}ska}}, \bibinfo {author}
  {\bibfnamefont {Y.}~\bibnamefont {Skourski}}, \bibinfo {author}
  {\bibfnamefont {D.}~\bibnamefont {Preusche}}, \bibinfo {author}
  {\bibfnamefont {M.}~\bibnamefont {Grifoni}}, \bibinfo {author} {\bibfnamefont
  {J.}~\bibnamefont {Wosnitza}},\ and\ \bibinfo {author} {\bibfnamefont
  {C.}~\bibnamefont {Strunk}},\ }\bibfield  {title} {\bibinfo {title} {{Direct
  Observation of Band-Gap Closure for a Semiconducting Carbon Nanotube in a
  Large Parallel Magnetic Field}},\ }\href
  {https://doi.org/10.1103/PhysRevLett.106.096802} {\bibfield  {journal}
  {\bibinfo  {journal} {Phys. Rev. Lett.}\ }\textbf {\bibinfo {volume} {106}},\
  \bibinfo {pages} {096802} (\bibinfo {year} {2011})}\BibitemShut {NoStop}%
\bibitem [{\citenamefont {Jespersen}\ \emph {et~al.}(2011)\citenamefont
  {Jespersen}, \citenamefont {Grove-Rasmussen}, \citenamefont {Paaske},
  \citenamefont {Muraki}, \citenamefont {Fujisawa}, \citenamefont
  {Nyg{\aa}rd},\ and\ \citenamefont {Flensberg}}]{Jespersen2011}%
  \BibitemOpen
  \bibfield  {author} {\bibinfo {author} {\bibfnamefont {T.~S.}\ \bibnamefont
  {Jespersen}}, \bibinfo {author} {\bibfnamefont {K.}~\bibnamefont
  {Grove-Rasmussen}}, \bibinfo {author} {\bibfnamefont {J.}~\bibnamefont
  {Paaske}}, \bibinfo {author} {\bibfnamefont {K.}~\bibnamefont {Muraki}},
  \bibinfo {author} {\bibfnamefont {T.}~\bibnamefont {Fujisawa}}, \bibinfo
  {author} {\bibfnamefont {J.}~\bibnamefont {Nyg{\aa}rd}},\ and\ \bibinfo
  {author} {\bibfnamefont {K.}~\bibnamefont {Flensberg}},\ }\bibfield  {title}
  {\bibinfo {title} {Gate-dependent spin--orbit coupling in multielectron
  carbon nanotubes},\ }\href {https://doi.org/10.1038/nphys1880} {\bibfield
  {journal} {\bibinfo  {journal} {Nature Physics}\ }\textbf {\bibinfo {volume}
  {7}},\ \bibinfo {pages} {348} (\bibinfo {year} {2011})}\BibitemShut {NoStop}%
\bibitem [{\citenamefont {Steele}\ \emph {et~al.}(2013)\citenamefont {Steele},
  \citenamefont {Pei}, \citenamefont {Laird}, \citenamefont {Jol},
  \citenamefont {Meerwaldt},\ and\ \citenamefont {Kouwenhoven}}]{Steele2013}%
  \BibitemOpen
  \bibfield  {author} {\bibinfo {author} {\bibfnamefont {G.~A.}\ \bibnamefont
  {Steele}}, \bibinfo {author} {\bibfnamefont {F.}~\bibnamefont {Pei}},
  \bibinfo {author} {\bibfnamefont {E.~A.}\ \bibnamefont {Laird}}, \bibinfo
  {author} {\bibfnamefont {J.~M.}\ \bibnamefont {Jol}}, \bibinfo {author}
  {\bibfnamefont {H.~B.}\ \bibnamefont {Meerwaldt}},\ and\ \bibinfo {author}
  {\bibfnamefont {L.~P.}\ \bibnamefont {Kouwenhoven}},\ }\bibfield  {title}
  {\bibinfo {title} {Large spin-orbit coupling in carbon nanotubes},\ }\href
  {https://doi.org/10.1038/ncomms2584} {\bibfield  {journal} {\bibinfo
  {journal} {Nature Communications}\ }\textbf {\bibinfo {volume} {4}},\
  \bibinfo {pages} {1573} (\bibinfo {year} {2013})}\BibitemShut {NoStop}%
\bibitem [{\citenamefont {Cao}\ \emph {et~al.}(2005)\citenamefont {Cao},
  \citenamefont {Wang},\ and\ \citenamefont {Dai}}]{Cao2005}%
  \BibitemOpen
  \bibfield  {author} {\bibinfo {author} {\bibfnamefont {J.}~\bibnamefont
  {Cao}}, \bibinfo {author} {\bibfnamefont {Q.}~\bibnamefont {Wang}},\ and\
  \bibinfo {author} {\bibfnamefont {H.}~\bibnamefont {Dai}},\ }\bibfield
  {title} {\bibinfo {title} {Electron transport in very clean, as-grown
  suspended carbon nanotubes},\ }\href@noop {} {\bibfield  {journal} {\bibinfo
  {journal} {Nat. Materials}\ }\textbf {\bibinfo {volume} {4}},\ \bibinfo
  {pages} {745} (\bibinfo {year} {2005})}\BibitemShut {NoStop}%
\bibitem [{\citenamefont {Wu}\ \emph {et~al.}(2010)\citenamefont {Wu},
  \citenamefont {Liu},\ and\ \citenamefont {Zhong}}]{Wu2010}%
  \BibitemOpen
  \bibfield  {author} {\bibinfo {author} {\bibfnamefont {C.~C.}\ \bibnamefont
  {Wu}}, \bibinfo {author} {\bibfnamefont {C.~H.}\ \bibnamefont {Liu}},\ and\
  \bibinfo {author} {\bibfnamefont {Z.}~\bibnamefont {Zhong}},\ }\bibfield
  {title} {\bibinfo {title} {{One-Step Direct Transfer of Pristine
  Single-Walled Carbon Nanotubes for Functional Nanoelectronics}},\ }\href
  {https://doi.org/10.1021/nl904260k} {\bibfield  {journal} {\bibinfo
  {journal} {Nano Letters}\ }\textbf {\bibinfo {volume} {10}},\ \bibinfo
  {pages} {1032} (\bibinfo {year} {2010})}\BibitemShut {NoStop}%
\bibitem [{\citenamefont {Pei}\ \emph {et~al.}(2012)\citenamefont {Pei},
  \citenamefont {Laird}, \citenamefont {Steele},\ and\ \citenamefont
  {Kouwenhoven}}]{Pei2012}%
  \BibitemOpen
  \bibfield  {author} {\bibinfo {author} {\bibfnamefont {F.}~\bibnamefont
  {Pei}}, \bibinfo {author} {\bibfnamefont {E.~A.}\ \bibnamefont {Laird}},
  \bibinfo {author} {\bibfnamefont {G.~A.}\ \bibnamefont {Steele}},\ and\
  \bibinfo {author} {\bibfnamefont {L.~P.}\ \bibnamefont {Kouwenhoven}},\
  }\bibfield  {title} {\bibinfo {title} {Valley--spin blockade and spin
  resonance in carbon nanotubes},\ }\href
  {https://doi.org/10.1038/nnano.2012.160} {\bibfield  {journal} {\bibinfo
  {journal} {Nature Nanotechnology}\ }\textbf {\bibinfo {volume} {7}},\
  \bibinfo {pages} {630} (\bibinfo {year} {2012})}\BibitemShut {NoStop}%
\bibitem [{\citenamefont {Ranjan}\ \emph {et~al.}(2015)\citenamefont {Ranjan},
  \citenamefont {Puebla-Hellmann}, \citenamefont {Jung}, \citenamefont
  {Hasler}, \citenamefont {Nunnenkamp}, \citenamefont {Muoth}, \citenamefont
  {Hierold}, \citenamefont {Wallraff},\ and\ \citenamefont
  {Sch{\"o}nenberger}}]{Ranjan2015}%
  \BibitemOpen
  \bibfield  {author} {\bibinfo {author} {\bibfnamefont {V.}~\bibnamefont
  {Ranjan}}, \bibinfo {author} {\bibfnamefont {G.}~\bibnamefont
  {Puebla-Hellmann}}, \bibinfo {author} {\bibfnamefont {M.}~\bibnamefont
  {Jung}}, \bibinfo {author} {\bibfnamefont {T.}~\bibnamefont {Hasler}},
  \bibinfo {author} {\bibfnamefont {A.}~\bibnamefont {Nunnenkamp}}, \bibinfo
  {author} {\bibfnamefont {M.}~\bibnamefont {Muoth}}, \bibinfo {author}
  {\bibfnamefont {C.}~\bibnamefont {Hierold}}, \bibinfo {author} {\bibfnamefont
  {A.}~\bibnamefont {Wallraff}},\ and\ \bibinfo {author} {\bibfnamefont
  {C.}~\bibnamefont {Sch{\"o}nenberger}},\ }\bibfield  {title} {\bibinfo
  {title} {Clean carbon nanotubes coupled to superconducting impedance-matching
  circuits},\ }\href {https://doi.org/10.1038/ncomms8165} {\bibfield  {journal}
  {\bibinfo  {journal} {Nature Communications}\ }\textbf {\bibinfo {volume}
  {6}},\ \bibinfo {pages} {7165} (\bibinfo {year} {2015})}\BibitemShut
  {NoStop}%
\bibitem [{\citenamefont {Waissman}\ \emph {et~al.}(2013)\citenamefont
  {Waissman}, \citenamefont {Honig}, \citenamefont {Pecker}, \citenamefont
  {Benyamini}, \citenamefont {Hamo},\ and\ \citenamefont
  {Ilani}}]{Waissman2013}%
  \BibitemOpen
  \bibfield  {author} {\bibinfo {author} {\bibfnamefont {J.}~\bibnamefont
  {Waissman}}, \bibinfo {author} {\bibfnamefont {M.}~\bibnamefont {Honig}},
  \bibinfo {author} {\bibfnamefont {S.}~\bibnamefont {Pecker}}, \bibinfo
  {author} {\bibfnamefont {A.}~\bibnamefont {Benyamini}}, \bibinfo {author}
  {\bibfnamefont {A.}~\bibnamefont {Hamo}},\ and\ \bibinfo {author}
  {\bibfnamefont {S.}~\bibnamefont {Ilani}},\ }\bibfield  {title} {\bibinfo
  {title} {Realization of pristine and locally tunable one-dimensional electron
  systems in carbon nanotubes},\ }\href
  {https://doi.org/10.1038/nnano.2013.143} {\bibfield  {journal} {\bibinfo
  {journal} {Nature Nanotechnology}\ }\textbf {\bibinfo {volume} {8}},\
  \bibinfo {pages} {569} (\bibinfo {year} {2013})}\BibitemShut {NoStop}%
\bibitem [{\citenamefont {Gramich}\ \emph {et~al.}(2015)\citenamefont
  {Gramich}, \citenamefont {Baumgartner}, \citenamefont {Muoth}, \citenamefont
  {Hierold},\ and\ \citenamefont {Schönenberger}}]{Gramich2015}%
  \BibitemOpen
  \bibfield  {author} {\bibinfo {author} {\bibfnamefont {J.}~\bibnamefont
  {Gramich}}, \bibinfo {author} {\bibfnamefont {A.}~\bibnamefont
  {Baumgartner}}, \bibinfo {author} {\bibfnamefont {M.}~\bibnamefont {Muoth}},
  \bibinfo {author} {\bibfnamefont {C.}~\bibnamefont {Hierold}},\ and\ \bibinfo
  {author} {\bibfnamefont {C.}~\bibnamefont {Schönenberger}},\ }\bibfield
  {title} {\bibinfo {title} {{Fork stamping of pristine carbon nanotubes onto
  ferromagnetic contacts for spin-valve devices}},\ }\href@noop {} {\bibfield
  {journal} {\bibinfo  {journal} {physica status solidi (b)}\ }\textbf
  {\bibinfo {volume} {252}},\ \bibinfo {pages} {2496} (\bibinfo {year}
  {2015})}\BibitemShut {NoStop}%
\bibitem [{\citenamefont {Blien}\ \emph {et~al.}(2018)\citenamefont {Blien},
  \citenamefont {Steger}, \citenamefont {Albang}, \citenamefont {Paradiso},\
  and\ \citenamefont {Hüttel}}]{Blien2018}%
  \BibitemOpen
  \bibfield  {author} {\bibinfo {author} {\bibfnamefont {S.}~\bibnamefont
  {Blien}}, \bibinfo {author} {\bibfnamefont {P.}~\bibnamefont {Steger}},
  \bibinfo {author} {\bibfnamefont {A.}~\bibnamefont {Albang}}, \bibinfo
  {author} {\bibfnamefont {N.}~\bibnamefont {Paradiso}},\ and\ \bibinfo
  {author} {\bibfnamefont {A.~K.}\ \bibnamefont {Hüttel}},\ }\bibfield
  {title} {\bibinfo {title} {{Quartz Tuning-Fork Based Carbon Nanotube Transfer
  into Quantum Device Geometries}},\ }\href@noop {} {\bibfield  {journal}
  {\bibinfo  {journal} {physica status solidi (b)}\ }\textbf {\bibinfo {volume}
  {255}},\ \bibinfo {pages} {1800118} (\bibinfo {year} {2018})}\BibitemShut
  {NoStop}%
\bibitem [{\citenamefont {Meyer}\ \emph {et~al.}(2007)\citenamefont {Meyer},
  \citenamefont {Geim}, \citenamefont {Katsnelson}, \citenamefont {Novoselov},
  \citenamefont {Booth},\ and\ \citenamefont {Roth}}]{Meyer2007}%
  \BibitemOpen
  \bibfield  {author} {\bibinfo {author} {\bibfnamefont {J.~C.}\ \bibnamefont
  {Meyer}}, \bibinfo {author} {\bibfnamefont {A.~K.}\ \bibnamefont {Geim}},
  \bibinfo {author} {\bibfnamefont {M.~I.}\ \bibnamefont {Katsnelson}},
  \bibinfo {author} {\bibfnamefont {K.~S.}\ \bibnamefont {Novoselov}}, \bibinfo
  {author} {\bibfnamefont {T.~J.}\ \bibnamefont {Booth}},\ and\ \bibinfo
  {author} {\bibfnamefont {S.}~\bibnamefont {Roth}},\ }\bibfield  {title}
  {\bibinfo {title} {The structure of suspended graphene sheets},\ }\href
  {https://doi.org/10.1038/nature05545} {\bibfield  {journal} {\bibinfo
  {journal} {Nature}\ }\textbf {\bibinfo {volume} {446}},\ \bibinfo {pages}
  {60} (\bibinfo {year} {2007})}\BibitemShut {NoStop}%
\bibitem [{\citenamefont {Du}\ \emph {et~al.}(2008)\citenamefont {Du},
  \citenamefont {Skachko}, \citenamefont {Barker},\ and\ \citenamefont
  {Andrei}}]{Du2008}%
  \BibitemOpen
  \bibfield  {author} {\bibinfo {author} {\bibfnamefont {X.}~\bibnamefont
  {Du}}, \bibinfo {author} {\bibfnamefont {I.}~\bibnamefont {Skachko}},
  \bibinfo {author} {\bibfnamefont {A.}~\bibnamefont {Barker}},\ and\ \bibinfo
  {author} {\bibfnamefont {E.~Y.}\ \bibnamefont {Andrei}},\ }\bibfield  {title}
  {\bibinfo {title} {Approaching ballistic transport in suspended graphene},\
  }\href {https://doi.org/10.1038/nnano.2008.199} {\bibfield  {journal}
  {\bibinfo  {journal} {Nature Nanotechnology}\ }\textbf {\bibinfo {volume}
  {3}},\ \bibinfo {pages} {491} (\bibinfo {year} {2008})}\BibitemShut {NoStop}%
\bibitem [{\citenamefont {Wang}\ \emph {et~al.}(2013)\citenamefont {Wang},
  \citenamefont {Meric}, \citenamefont {Huang}, \citenamefont {Gao},
  \citenamefont {Gao}, \citenamefont {Tran}, \citenamefont {Taniguchi},
  \citenamefont {Watanabe}, \citenamefont {Campos}, \citenamefont {Muller},
  \citenamefont {Guo}, \citenamefont {Kim}, \citenamefont {Hone}, \citenamefont
  {Shepard},\ and\ \citenamefont {Dean}}]{Wang614}%
  \BibitemOpen
  \bibfield  {author} {\bibinfo {author} {\bibfnamefont {L.}~\bibnamefont
  {Wang}}, \bibinfo {author} {\bibfnamefont {I.}~\bibnamefont {Meric}},
  \bibinfo {author} {\bibfnamefont {P.~Y.}\ \bibnamefont {Huang}}, \bibinfo
  {author} {\bibfnamefont {Q.}~\bibnamefont {Gao}}, \bibinfo {author}
  {\bibfnamefont {Y.}~\bibnamefont {Gao}}, \bibinfo {author} {\bibfnamefont
  {H.}~\bibnamefont {Tran}}, \bibinfo {author} {\bibfnamefont {T.}~\bibnamefont
  {Taniguchi}}, \bibinfo {author} {\bibfnamefont {K.}~\bibnamefont {Watanabe}},
  \bibinfo {author} {\bibfnamefont {L.~M.}\ \bibnamefont {Campos}}, \bibinfo
  {author} {\bibfnamefont {D.~A.}\ \bibnamefont {Muller}}, \bibinfo {author}
  {\bibfnamefont {J.}~\bibnamefont {Guo}}, \bibinfo {author} {\bibfnamefont
  {P.}~\bibnamefont {Kim}}, \bibinfo {author} {\bibfnamefont {J.}~\bibnamefont
  {Hone}}, \bibinfo {author} {\bibfnamefont {K.~L.}\ \bibnamefont {Shepard}},\
  and\ \bibinfo {author} {\bibfnamefont {C.~R.}\ \bibnamefont {Dean}},\
  }\bibfield  {title} {\bibinfo {title} {{One-Dimensional Electrical Contact to
  a Two-Dimensional Material}},\ }\href@noop {} {\bibfield  {journal} {\bibinfo
   {journal} {Science}\ }\textbf {\bibinfo {volume} {342}},\ \bibinfo {pages}
  {614} (\bibinfo {year} {2013})}\BibitemShut {NoStop}%
\bibitem [{\citenamefont {Geim}\ and\ \citenamefont
  {Grigorieva}(2013)}]{Geim2013}%
  \BibitemOpen
  \bibfield  {author} {\bibinfo {author} {\bibfnamefont {A.~K.}\ \bibnamefont
  {Geim}}\ and\ \bibinfo {author} {\bibfnamefont {I.~V.}\ \bibnamefont
  {Grigorieva}},\ }\bibfield  {title} {\bibinfo {title} {{Van der Waals
  heterostructures}},\ }\href {https://doi.org/10.1038/nature12385} {\bibfield
  {journal} {\bibinfo  {journal} {Nature}\ }\textbf {\bibinfo {volume} {499}},\
  \bibinfo {pages} {419} (\bibinfo {year} {2013})}\BibitemShut {NoStop}%
\bibitem [{\citenamefont {Novoselov}\ \emph {et~al.}(2016)\citenamefont
  {Novoselov}, \citenamefont {Mishchenko}, \citenamefont {Carvalho},\ and\
  \citenamefont {Castro~Neto}}]{Novoselovaac9439}%
  \BibitemOpen
  \bibfield  {author} {\bibinfo {author} {\bibfnamefont {K.~S.}\ \bibnamefont
  {Novoselov}}, \bibinfo {author} {\bibfnamefont {A.}~\bibnamefont
  {Mishchenko}}, \bibinfo {author} {\bibfnamefont {A.}~\bibnamefont
  {Carvalho}},\ and\ \bibinfo {author} {\bibfnamefont {A.~H.}\ \bibnamefont
  {Castro~Neto}},\ }\bibfield  {title} {\bibinfo {title} {{2D materials and van
  der Waals heterostructures}},\ }\href@noop {} {\bibfield  {journal} {\bibinfo
   {journal} {Science}\ }\textbf {\bibinfo {volume} {353}} (\bibinfo {year}
  {2016})}\BibitemShut {NoStop}%
\bibitem [{\citenamefont {Paradiso}\ \emph {et~al.}(2019)\citenamefont
  {Paradiso}, \citenamefont {Nguyen}, \citenamefont {Kloss},\ and\
  \citenamefont {Strunk}}]{Paradiso_2DMat_2019}%
  \BibitemOpen
  \bibfield  {author} {\bibinfo {author} {\bibfnamefont {N.}~\bibnamefont
  {Paradiso}}, \bibinfo {author} {\bibfnamefont {A.-T.}\ \bibnamefont
  {Nguyen}}, \bibinfo {author} {\bibfnamefont {K.~E.}\ \bibnamefont {Kloss}},\
  and\ \bibinfo {author} {\bibfnamefont {C.}~\bibnamefont {Strunk}},\
  }\bibfield  {title} {\bibinfo {title} {{Phase slip lines in superconducting
  few-layer NbSe$_2$ crystals}},\ }\href
  {https://doi.org/10.1088/2053-1583/ab0bcc} {\bibfield  {journal} {\bibinfo
  {journal} {2D Materials}\ }\textbf {\bibinfo {volume} {6}},\ \bibinfo {pages}
  {025039} (\bibinfo {year} {2019})}\BibitemShut {NoStop}%
\bibitem [{\citenamefont {Castellanos-Gomez}\ \emph {et~al.}(2014)\citenamefont
  {Castellanos-Gomez}, \citenamefont {Buscema}, \citenamefont {Molenaar},
  \citenamefont {Singh}, \citenamefont {Janssen}, \citenamefont {van~der
  Zant},\ and\ \citenamefont {Steele}}]{Castellanos_Gomez_2014}%
  \BibitemOpen
  \bibfield  {author} {\bibinfo {author} {\bibfnamefont {A.}~\bibnamefont
  {Castellanos-Gomez}}, \bibinfo {author} {\bibfnamefont {M.}~\bibnamefont
  {Buscema}}, \bibinfo {author} {\bibfnamefont {R.}~\bibnamefont {Molenaar}},
  \bibinfo {author} {\bibfnamefont {V.}~\bibnamefont {Singh}}, \bibinfo
  {author} {\bibfnamefont {L.}~\bibnamefont {Janssen}}, \bibinfo {author}
  {\bibfnamefont {H.~S.~J.}\ \bibnamefont {van~der Zant}},\ and\ \bibinfo
  {author} {\bibfnamefont {G.~A.}\ \bibnamefont {Steele}},\ }\bibfield  {title}
  {\bibinfo {title} {Deterministic transfer of two-dimensional materials by
  all-dry viscoelastic stamping},\ }\href
  {https://doi.org/10.1088/2053-1583/1/1/011002} {\bibfield  {journal}
  {\bibinfo  {journal} {2D Materials}\ }\textbf {\bibinfo {volume} {1}},\
  \bibinfo {pages} {011002} (\bibinfo {year} {2014})}\BibitemShut {NoStop}%
\bibitem [{\citenamefont {Overweg}\ \emph
  {et~al.}(2018{\natexlab{a}})\citenamefont {Overweg}, \citenamefont {Knothe},
  \citenamefont {Fabian}, \citenamefont {Linhart}, \citenamefont {Rickhaus},
  \citenamefont {Wernli}, \citenamefont {Watanabe}, \citenamefont {Taniguchi},
  \citenamefont {S\'anchez}, \citenamefont {Burgd\"orfer}, \citenamefont
  {Libisch}, \citenamefont {Fal'ko}, \citenamefont {Ensslin},\ and\
  \citenamefont {Ihn}}]{Overweg2018PRL}%
  \BibitemOpen
  \bibfield  {author} {\bibinfo {author} {\bibfnamefont {H.}~\bibnamefont
  {Overweg}}, \bibinfo {author} {\bibfnamefont {A.}~\bibnamefont {Knothe}},
  \bibinfo {author} {\bibfnamefont {T.}~\bibnamefont {Fabian}}, \bibinfo
  {author} {\bibfnamefont {L.}~\bibnamefont {Linhart}}, \bibinfo {author}
  {\bibfnamefont {P.}~\bibnamefont {Rickhaus}}, \bibinfo {author}
  {\bibfnamefont {L.}~\bibnamefont {Wernli}}, \bibinfo {author} {\bibfnamefont
  {K.}~\bibnamefont {Watanabe}}, \bibinfo {author} {\bibfnamefont
  {T.}~\bibnamefont {Taniguchi}}, \bibinfo {author} {\bibfnamefont
  {D.}~\bibnamefont {S\'anchez}}, \bibinfo {author} {\bibfnamefont
  {J.}~\bibnamefont {Burgd\"orfer}}, \bibinfo {author} {\bibfnamefont
  {F.}~\bibnamefont {Libisch}}, \bibinfo {author} {\bibfnamefont {V.~I.}\
  \bibnamefont {Fal'ko}}, \bibinfo {author} {\bibfnamefont {K.}~\bibnamefont
  {Ensslin}},\ and\ \bibinfo {author} {\bibfnamefont {T.}~\bibnamefont {Ihn}},\
  }\bibfield  {title} {\bibinfo {title} {{Topologically Nontrivial Valley
  States in Bilayer Graphene Quantum Point Contacts}},\ }\href
  {https://doi.org/10.1103/PhysRevLett.121.257702} {\bibfield  {journal}
  {\bibinfo  {journal} {Phys. Rev. Lett.}\ }\textbf {\bibinfo {volume} {121}},\
  \bibinfo {pages} {257702} (\bibinfo {year} {2018}{\natexlab{a}})}\BibitemShut
  {NoStop}%
\bibitem [{\citenamefont {Overweg}\ \emph
  {et~al.}(2018{\natexlab{b}})\citenamefont {Overweg}, \citenamefont
  {Eggimann}, \citenamefont {Chen}, \citenamefont {Slizovskiy}, \citenamefont
  {Eich}, \citenamefont {Pisoni}, \citenamefont {Lee}, \citenamefont
  {Rickhaus}, \citenamefont {Watanabe}, \citenamefont {Taniguchi},
  \citenamefont {Fal'ko}, \citenamefont {Ihn},\ and\ \citenamefont
  {Ensslin}}]{Overweg2018NL}%
  \BibitemOpen
  \bibfield  {author} {\bibinfo {author} {\bibfnamefont {H.}~\bibnamefont
  {Overweg}}, \bibinfo {author} {\bibfnamefont {H.}~\bibnamefont {Eggimann}},
  \bibinfo {author} {\bibfnamefont {X.}~\bibnamefont {Chen}}, \bibinfo {author}
  {\bibfnamefont {S.}~\bibnamefont {Slizovskiy}}, \bibinfo {author}
  {\bibfnamefont {M.}~\bibnamefont {Eich}}, \bibinfo {author} {\bibfnamefont
  {R.}~\bibnamefont {Pisoni}}, \bibinfo {author} {\bibfnamefont
  {Y.}~\bibnamefont {Lee}}, \bibinfo {author} {\bibfnamefont {P.}~\bibnamefont
  {Rickhaus}}, \bibinfo {author} {\bibfnamefont {K.}~\bibnamefont {Watanabe}},
  \bibinfo {author} {\bibfnamefont {T.}~\bibnamefont {Taniguchi}}, \bibinfo
  {author} {\bibfnamefont {V.}~\bibnamefont {Fal'ko}}, \bibinfo {author}
  {\bibfnamefont {T.}~\bibnamefont {Ihn}},\ and\ \bibinfo {author}
  {\bibfnamefont {K.}~\bibnamefont {Ensslin}},\ }\bibfield  {title} {\bibinfo
  {title} {{Electrostatically Induced Quantum Point Contacts in Bilayer
  Graphene}},\ }\href {https://doi.org/10.1021/acs.nanolett.7b04666} {\bibfield
   {journal} {\bibinfo  {journal} {Nano Letters}\ }\textbf {\bibinfo {volume}
  {18}},\ \bibinfo {pages} {553} (\bibinfo {year}
  {2018}{\natexlab{b}})}\BibitemShut {NoStop}%
\bibitem [{\citenamefont {Huang}\ \emph {et~al.}(2015)\citenamefont {Huang},
  \citenamefont {Pan}, \citenamefont {Tran}, \citenamefont {Cheng},
  \citenamefont {Watanabe}, \citenamefont {Taniguchi}, \citenamefont {Lau},\
  and\ \citenamefont {Bockrath}}]{HuangCNTedge}%
  \BibitemOpen
  \bibfield  {author} {\bibinfo {author} {\bibfnamefont {J.-W.}\ \bibnamefont
  {Huang}}, \bibinfo {author} {\bibfnamefont {C.}~\bibnamefont {Pan}}, \bibinfo
  {author} {\bibfnamefont {S.}~\bibnamefont {Tran}}, \bibinfo {author}
  {\bibfnamefont {B.}~\bibnamefont {Cheng}}, \bibinfo {author} {\bibfnamefont
  {K.}~\bibnamefont {Watanabe}}, \bibinfo {author} {\bibfnamefont
  {T.}~\bibnamefont {Taniguchi}}, \bibinfo {author} {\bibfnamefont {C.~N.}\
  \bibnamefont {Lau}},\ and\ \bibinfo {author} {\bibfnamefont {M.}~\bibnamefont
  {Bockrath}},\ }\bibfield  {title} {\bibinfo {title} {{Superior Current
  Carrying Capacity of Boron Nitride Encapsulated Carbon Nanotubes with
  Zero-Dimensional Contacts}},\ }\href@noop {} {\bibfield  {journal} {\bibinfo
  {journal} {Nano Letters}\ }\textbf {\bibinfo {volume} {15}},\ \bibinfo
  {pages} {6836} (\bibinfo {year} {2015})}\BibitemShut {NoStop}%
\bibitem [{Note1()}]{Note1}%
  \BibitemOpen
  \bibinfo {note} {See Supplementary Material for further
  information}\BibitemShut {NoStop}%
\bibitem [{\citenamefont {Kasumov}\ \emph {et~al.}(1999)\citenamefont
  {Kasumov}, \citenamefont {Deblock}, \citenamefont {Kociak}, \citenamefont
  {Reulet}, \citenamefont {Bouchiat}, \citenamefont {Khodos}, \citenamefont
  {Gorbatov}, \citenamefont {Volkov}, \citenamefont {Journet},\ and\
  \citenamefont {Burghard}}]{Kasumov1508}%
  \BibitemOpen
  \bibfield  {author} {\bibinfo {author} {\bibfnamefont {A.~Y.}\ \bibnamefont
  {Kasumov}}, \bibinfo {author} {\bibfnamefont {R.}~\bibnamefont {Deblock}},
  \bibinfo {author} {\bibfnamefont {M.}~\bibnamefont {Kociak}}, \bibinfo
  {author} {\bibfnamefont {B.}~\bibnamefont {Reulet}}, \bibinfo {author}
  {\bibfnamefont {H.}~\bibnamefont {Bouchiat}}, \bibinfo {author}
  {\bibfnamefont {I.~I.}\ \bibnamefont {Khodos}}, \bibinfo {author}
  {\bibfnamefont {Y.~B.}\ \bibnamefont {Gorbatov}}, \bibinfo {author}
  {\bibfnamefont {V.~T.}\ \bibnamefont {Volkov}}, \bibinfo {author}
  {\bibfnamefont {C.}~\bibnamefont {Journet}},\ and\ \bibinfo {author}
  {\bibfnamefont {M.}~\bibnamefont {Burghard}},\ }\bibfield  {title} {\bibinfo
  {title} {{Supercurrents Through Single-Walled Carbon Nanotubes}},\
  }\href@noop {} {\bibfield  {journal} {\bibinfo  {journal} {Science}\ }\textbf
  {\bibinfo {volume} {284}},\ \bibinfo {pages} {1508} (\bibinfo {year}
  {1999})}\BibitemShut {NoStop}%
\bibitem [{\citenamefont {Kasumov}\ \emph {et~al.}(2003)\citenamefont
  {Kasumov}, \citenamefont {Kociak}, \citenamefont {Ferrier}, \citenamefont
  {Deblock}, \citenamefont {Gu\'eron}, \citenamefont {Reulet}, \citenamefont
  {Khodos}, \citenamefont {St\'ephan},\ and\ \citenamefont
  {Bouchiat}}]{KasumovPRB2003}%
  \BibitemOpen
  \bibfield  {author} {\bibinfo {author} {\bibfnamefont {A.}~\bibnamefont
  {Kasumov}}, \bibinfo {author} {\bibfnamefont {M.}~\bibnamefont {Kociak}},
  \bibinfo {author} {\bibfnamefont {M.}~\bibnamefont {Ferrier}}, \bibinfo
  {author} {\bibfnamefont {R.}~\bibnamefont {Deblock}}, \bibinfo {author}
  {\bibfnamefont {S.}~\bibnamefont {Gu\'eron}}, \bibinfo {author}
  {\bibfnamefont {B.}~\bibnamefont {Reulet}}, \bibinfo {author} {\bibfnamefont
  {I.}~\bibnamefont {Khodos}}, \bibinfo {author} {\bibfnamefont
  {O.}~\bibnamefont {St\'ephan}},\ and\ \bibinfo {author} {\bibfnamefont
  {H.}~\bibnamefont {Bouchiat}},\ }\bibfield  {title} {\bibinfo {title}
  {{Quantum transport through carbon nanotubes: Proximity-induced and intrinsic
  superconductivity}},\ }\href {https://doi.org/10.1103/PhysRevB.68.214521}
  {\bibfield  {journal} {\bibinfo  {journal} {Phys. Rev. B}\ }\textbf {\bibinfo
  {volume} {68}},\ \bibinfo {pages} {214521} (\bibinfo {year}
  {2003})}\BibitemShut {NoStop}%
\bibitem [{\citenamefont {Ferrier}\ \emph {et~al.}(2004)\citenamefont
  {Ferrier}, \citenamefont {{De Martino}}, \citenamefont {Kasumov},
  \citenamefont {Guéron}, \citenamefont {Kociak}, \citenamefont {Egger},\ and\
  \citenamefont {Bouchiat}}]{Ferrier2004}%
  \BibitemOpen
  \bibfield  {author} {\bibinfo {author} {\bibfnamefont {M.}~\bibnamefont
  {Ferrier}}, \bibinfo {author} {\bibfnamefont {A.}~\bibnamefont {{De
  Martino}}}, \bibinfo {author} {\bibfnamefont {A.}~\bibnamefont {Kasumov}},
  \bibinfo {author} {\bibfnamefont {S.}~\bibnamefont {Guéron}}, \bibinfo
  {author} {\bibfnamefont {M.}~\bibnamefont {Kociak}}, \bibinfo {author}
  {\bibfnamefont {R.}~\bibnamefont {Egger}},\ and\ \bibinfo {author}
  {\bibfnamefont {H.}~\bibnamefont {Bouchiat}},\ }\bibfield  {title} {\bibinfo
  {title} {Superconductivity in ropes of carbon nanotubes},\ }\href@noop {}
  {\bibfield  {journal} {\bibinfo  {journal} {Solid State Communications}\
  }\textbf {\bibinfo {volume} {131}},\ \bibinfo {pages} {615 } (\bibinfo {year}
  {2004})}\BibitemShut {NoStop}%
\bibitem [{\citenamefont {Arutyunov}\ \emph {et~al.}(2008)\citenamefont
  {Arutyunov}, \citenamefont {Golubev},\ and\ \citenamefont
  {Zaikin}}]{ARUTYUNOV20081}%
  \BibitemOpen
  \bibfield  {author} {\bibinfo {author} {\bibfnamefont {K.}~\bibnamefont
  {Arutyunov}}, \bibinfo {author} {\bibfnamefont {D.}~\bibnamefont {Golubev}},\
  and\ \bibinfo {author} {\bibfnamefont {A.}~\bibnamefont {Zaikin}},\
  }\bibfield  {title} {\bibinfo {title} {Superconductivity in one dimension},\
  }\href {https://doi.org/https://doi.org/10.1016/j.physrep.2008.04.009}
  {\bibfield  {journal} {\bibinfo  {journal} {Physics Reports}\ }\textbf
  {\bibinfo {volume} {464}},\ \bibinfo {pages} {1 } (\bibinfo {year}
  {2008})}\BibitemShut {NoStop}%
\bibitem [{\citenamefont {Ivlev}\ and\ \citenamefont
  {Kopnin}(1984)}]{IvlevKopninAdv1984}%
  \BibitemOpen
  \bibfield  {author} {\bibinfo {author} {\bibfnamefont {B.}~\bibnamefont
  {Ivlev}}\ and\ \bibinfo {author} {\bibfnamefont {N.}~\bibnamefont {Kopnin}},\
  }\bibfield  {title} {\bibinfo {title} {Electric currents and resistive states
  in thin superconductors},\ }\href {https://doi.org/10.1080/00018738400101641}
  {\bibfield  {journal} {\bibinfo  {journal} {Advances in Physics}\ }\textbf
  {\bibinfo {volume} {33}},\ \bibinfo {pages} {47} (\bibinfo {year}
  {1984})}\BibitemShut {NoStop}%
\bibitem [{\citenamefont {Bezryadin}(2013)}]{Bezryadin2013}%
  \BibitemOpen
  \bibfield  {author} {\bibinfo {author} {\bibfnamefont {A.}~\bibnamefont
  {Bezryadin}},\ }\href@noop {} {\emph {\bibinfo {title} {{Superconductivity in
  Nanowires: Fabrication and Quantum Transport}}}}\ (\bibinfo  {publisher}
  {WILEY-VCH Verlag},\ \bibinfo {address} {Weinheim, Germany},\ \bibinfo {year}
  {2013})\BibitemShut {NoStop}%
\bibitem [{\citenamefont {Michotte}\ \emph {et~al.}(2004)\citenamefont
  {Michotte}, \citenamefont {M\'at\'efi-Tempfli}, \citenamefont {Piraux},
  \citenamefont {Vodolazov},\ and\ \citenamefont {Peeters}}]{Michotte2004}%
  \BibitemOpen
  \bibfield  {author} {\bibinfo {author} {\bibfnamefont {S.}~\bibnamefont
  {Michotte}}, \bibinfo {author} {\bibfnamefont {S.}~\bibnamefont
  {M\'at\'efi-Tempfli}}, \bibinfo {author} {\bibfnamefont {L.}~\bibnamefont
  {Piraux}}, \bibinfo {author} {\bibfnamefont {D.~Y.}\ \bibnamefont
  {Vodolazov}},\ and\ \bibinfo {author} {\bibfnamefont {F.~M.}\ \bibnamefont
  {Peeters}},\ }\bibfield  {title} {\bibinfo {title} {{Condition for the
  occurrence of phase slip centers in superconducting nanowires under applied
  current or voltage}},\ }\href {https://doi.org/10.1103/PhysRevB.69.094512}
  {\bibfield  {journal} {\bibinfo  {journal} {Phys. Rev. B}\ }\textbf {\bibinfo
  {volume} {69}},\ \bibinfo {pages} {094512} (\bibinfo {year}
  {2004})}\BibitemShut {NoStop}%
\bibitem [{\citenamefont {Vodolazov}(2007)}]{Vodolazov2007}%
  \BibitemOpen
  \bibfield  {author} {\bibinfo {author} {\bibfnamefont {D.~Y.}\ \bibnamefont
  {Vodolazov}},\ }\bibfield  {title} {\bibinfo {title} {{Negative
  magnetoresistance and phase slip process in superconducting nanowires}},\
  }\href {https://doi.org/10.1103/PhysRevB.75.184517} {\bibfield  {journal}
  {\bibinfo  {journal} {Phys. Rev. B}\ }\textbf {\bibinfo {volume} {75}},\
  \bibinfo {pages} {184517} (\bibinfo {year} {2007})}\BibitemShut {NoStop}%
\bibitem [{\citenamefont {Tran}\ \emph {et~al.}(2020)\citenamefont {Tran},
  \citenamefont {Sell},\ and\ \citenamefont {Williams}}]{Tran}%
  \BibitemOpen
  \bibfield  {author} {\bibinfo {author} {\bibfnamefont {S.}~\bibnamefont
  {Tran}}, \bibinfo {author} {\bibfnamefont {J.}~\bibnamefont {Sell}},\ and\
  \bibinfo {author} {\bibfnamefont {J.~R.}\ \bibnamefont {Williams}},\
  }\bibfield  {title} {\bibinfo {title} {{Dynamical Josephson effects in
  $\mathrm{Nb}{\mathrm{Se}}_{2}$}},\ }\href
  {https://doi.org/10.1103/PhysRevResearch.2.043204} {\bibfield  {journal}
  {\bibinfo  {journal} {Phys. Rev. Research}\ }\textbf {\bibinfo {volume}
  {2}},\ \bibinfo {pages} {043204} (\bibinfo {year} {2020})}\BibitemShut
  {NoStop}%
\bibitem [{\citenamefont {Saito}\ \emph {et~al.}(2020)\citenamefont {Saito},
  \citenamefont {Itahashi}, \citenamefont {Nojima},\ and\ \citenamefont
  {Iwasa}}]{SaitoPRMat}%
  \BibitemOpen
  \bibfield  {author} {\bibinfo {author} {\bibfnamefont {Y.}~\bibnamefont
  {Saito}}, \bibinfo {author} {\bibfnamefont {Y.~M.}\ \bibnamefont {Itahashi}},
  \bibinfo {author} {\bibfnamefont {T.}~\bibnamefont {Nojima}},\ and\ \bibinfo
  {author} {\bibfnamefont {Y.}~\bibnamefont {Iwasa}},\ }\bibfield  {title}
  {\bibinfo {title} {{Dynamical vortex phase diagram of two-dimensional
  superconductivity in gated $\mathrm{Mo}{\mathrm{S}}_{2}$}},\ }\href
  {https://doi.org/10.1103/PhysRevMaterials.4.074003} {\bibfield  {journal}
  {\bibinfo  {journal} {Phys. Rev. Materials}\ }\textbf {\bibinfo {volume}
  {4}},\ \bibinfo {pages} {074003} (\bibinfo {year} {2020})}\BibitemShut
  {NoStop}%
\bibitem [{\citenamefont {Moriya}\ \emph {et~al.}(2020)\citenamefont {Moriya},
  \citenamefont {Yabuki},\ and\ \citenamefont {Machida}}]{Moriya2020}%
  \BibitemOpen
  \bibfield  {author} {\bibinfo {author} {\bibfnamefont {R.}~\bibnamefont
  {Moriya}}, \bibinfo {author} {\bibfnamefont {N.}~\bibnamefont {Yabuki}},\
  and\ \bibinfo {author} {\bibfnamefont {T.}~\bibnamefont {Machida}},\
  }\bibfield  {title} {\bibinfo {title} {{Superconducting proximity effect in a
  $\mathrm{Nb}{\mathrm{Se}}_{2}/\text{graphene}$ van der Waals junction}},\
  }\href {https://doi.org/10.1103/PhysRevB.101.054503} {\bibfield  {journal}
  {\bibinfo  {journal} {Phys. Rev. B}\ }\textbf {\bibinfo {volume} {101}},\
  \bibinfo {pages} {054503} (\bibinfo {year} {2020})}\BibitemShut {NoStop}%
\bibitem [{\citenamefont {Li}\ \emph {et~al.}(2020)\citenamefont {Li},
  \citenamefont {Leng}, \citenamefont {Fu}, \citenamefont {Watanabe},
  \citenamefont {Taniguchi}, \citenamefont {Liu}, \citenamefont {Liu},\ and\
  \citenamefont {Zhu}}]{Li2020}%
  \BibitemOpen
  \bibfield  {author} {\bibinfo {author} {\bibfnamefont {J.}~\bibnamefont
  {Li}}, \bibinfo {author} {\bibfnamefont {H.-B.}\ \bibnamefont {Leng}},
  \bibinfo {author} {\bibfnamefont {H.}~\bibnamefont {Fu}}, \bibinfo {author}
  {\bibfnamefont {K.}~\bibnamefont {Watanabe}}, \bibinfo {author}
  {\bibfnamefont {T.}~\bibnamefont {Taniguchi}}, \bibinfo {author}
  {\bibfnamefont {X.}~\bibnamefont {Liu}}, \bibinfo {author} {\bibfnamefont
  {C.-X.}\ \bibnamefont {Liu}},\ and\ \bibinfo {author} {\bibfnamefont
  {J.}~\bibnamefont {Zhu}},\ }\bibfield  {title} {\bibinfo {title}
  {{Superconducting proximity effect in a transparent van der Waals
  superconductor-metal junction}},\ }\href
  {https://doi.org/10.1103/PhysRevB.101.195405} {\bibfield  {journal} {\bibinfo
   {journal} {Phys. Rev. B}\ }\textbf {\bibinfo {volume} {101}},\ \bibinfo
  {pages} {195405} (\bibinfo {year} {2020})}\BibitemShut {NoStop}%
\bibitem [{Note2()}]{Note2}%
  \BibitemOpen
  \bibinfo {note} {The conductance measurements with in-plane magnetic field
  (Fig.~\ref {fig:allcolor}(d,e,f)) were performed in a second cool-down,
  necessary to rotate the sample.}\BibitemShut {Stop}%
\bibitem [{\citenamefont {Kociak}\ \emph {et~al.}(2001)\citenamefont {Kociak},
  \citenamefont {Kasumov}, \citenamefont {Gu\'eron}, \citenamefont {Reulet},
  \citenamefont {Khodos}, \citenamefont {Gorbatov}, \citenamefont {Volkov},
  \citenamefont {Vaccarini},\ and\ \citenamefont {Bouchiat}}]{Kociak2001}%
  \BibitemOpen
  \bibfield  {author} {\bibinfo {author} {\bibfnamefont {M.}~\bibnamefont
  {Kociak}}, \bibinfo {author} {\bibfnamefont {A.~Y.}\ \bibnamefont {Kasumov}},
  \bibinfo {author} {\bibfnamefont {S.}~\bibnamefont {Gu\'eron}}, \bibinfo
  {author} {\bibfnamefont {B.}~\bibnamefont {Reulet}}, \bibinfo {author}
  {\bibfnamefont {I.~I.}\ \bibnamefont {Khodos}}, \bibinfo {author}
  {\bibfnamefont {Y.~B.}\ \bibnamefont {Gorbatov}}, \bibinfo {author}
  {\bibfnamefont {V.~T.}\ \bibnamefont {Volkov}}, \bibinfo {author}
  {\bibfnamefont {L.}~\bibnamefont {Vaccarini}},\ and\ \bibinfo {author}
  {\bibfnamefont {H.}~\bibnamefont {Bouchiat}},\ }\bibfield  {title} {\bibinfo
  {title} {{Superconductivity in Ropes of Single-Walled Carbon Nanotubes}},\
  }\href {https://doi.org/10.1103/PhysRevLett.86.2416} {\bibfield  {journal}
  {\bibinfo  {journal} {Phys. Rev. Lett.}\ }\textbf {\bibinfo {volume} {86}},\
  \bibinfo {pages} {2416} (\bibinfo {year} {2001})}\BibitemShut {NoStop}%
\bibitem [{\citenamefont {Tang}\ \emph {et~al.}(2001)\citenamefont {Tang},
  \citenamefont {Zhang}, \citenamefont {Wang}, \citenamefont {Zhang},
  \citenamefont {Wen}, \citenamefont {Li}, \citenamefont {Wang}, \citenamefont
  {Chan},\ and\ \citenamefont {Sheng}}]{Tang2462}%
  \BibitemOpen
  \bibfield  {author} {\bibinfo {author} {\bibfnamefont {Z.~K.}\ \bibnamefont
  {Tang}}, \bibinfo {author} {\bibfnamefont {L.}~\bibnamefont {Zhang}},
  \bibinfo {author} {\bibfnamefont {N.}~\bibnamefont {Wang}}, \bibinfo {author}
  {\bibfnamefont {X.~X.}\ \bibnamefont {Zhang}}, \bibinfo {author}
  {\bibfnamefont {G.~H.}\ \bibnamefont {Wen}}, \bibinfo {author} {\bibfnamefont
  {G.~D.}\ \bibnamefont {Li}}, \bibinfo {author} {\bibfnamefont {J.~N.}\
  \bibnamefont {Wang}}, \bibinfo {author} {\bibfnamefont {C.~T.}\ \bibnamefont
  {Chan}},\ and\ \bibinfo {author} {\bibfnamefont {P.}~\bibnamefont {Sheng}},\
  }\bibfield  {title} {\bibinfo {title} {{Superconductivity in 4 Angstrom
  Single-Walled Carbon Nanotubes}},\ }\href
  {https://doi.org/10.1126/science.1060470} {\bibfield  {journal} {\bibinfo
  {journal} {Science}\ }\textbf {\bibinfo {volume} {292}},\ \bibinfo {pages}
  {2462} (\bibinfo {year} {2001})}\BibitemShut {NoStop}%
\bibitem [{\citenamefont {Takesue}\ \emph {et~al.}(2006)\citenamefont
  {Takesue}, \citenamefont {Haruyama}, \citenamefont {Kobayashi}, \citenamefont
  {Chiashi}, \citenamefont {Maruyama}, \citenamefont {Sugai},\ and\
  \citenamefont {Shinohara}}]{Takesue2006}%
  \BibitemOpen
  \bibfield  {author} {\bibinfo {author} {\bibfnamefont {I.}~\bibnamefont
  {Takesue}}, \bibinfo {author} {\bibfnamefont {J.}~\bibnamefont {Haruyama}},
  \bibinfo {author} {\bibfnamefont {N.}~\bibnamefont {Kobayashi}}, \bibinfo
  {author} {\bibfnamefont {S.}~\bibnamefont {Chiashi}}, \bibinfo {author}
  {\bibfnamefont {S.}~\bibnamefont {Maruyama}}, \bibinfo {author}
  {\bibfnamefont {T.}~\bibnamefont {Sugai}},\ and\ \bibinfo {author}
  {\bibfnamefont {H.}~\bibnamefont {Shinohara}},\ }\bibfield  {title} {\bibinfo
  {title} {{Superconductivity in Entirely End-Bonded Multiwalled Carbon
  Nanotubes}},\ }\href {https://doi.org/10.1103/PhysRevLett.96.057001}
  {\bibfield  {journal} {\bibinfo  {journal} {Phys. Rev. Lett.}\ }\textbf
  {\bibinfo {volume} {96}},\ \bibinfo {pages} {057001} (\bibinfo {year}
  {2006})}\BibitemShut {NoStop}%
\bibitem [{\citenamefont {Shi}\ \emph {et~al.}(2012)\citenamefont {Shi},
  \citenamefont {Wang}, \citenamefont {Zhang}, \citenamefont {Zheng},
  \citenamefont {Ieong}, \citenamefont {He}, \citenamefont {Lortz},
  \citenamefont {Cai}, \citenamefont {Wang}, \citenamefont {Zhang},
  \citenamefont {Zhang}, \citenamefont {Tang}, \citenamefont {Sheng},
  \citenamefont {Muramatsu}, \citenamefont {Kim}, \citenamefont {Endo},
  \citenamefont {Araujo},\ and\ \citenamefont {Dresselhaus}}]{Shi2012}%
  \BibitemOpen
  \bibfield  {author} {\bibinfo {author} {\bibfnamefont {W.}~\bibnamefont
  {Shi}}, \bibinfo {author} {\bibfnamefont {Z.}~\bibnamefont {Wang}}, \bibinfo
  {author} {\bibfnamefont {Q.}~\bibnamefont {Zhang}}, \bibinfo {author}
  {\bibfnamefont {Y.}~\bibnamefont {Zheng}}, \bibinfo {author} {\bibfnamefont
  {C.}~\bibnamefont {Ieong}}, \bibinfo {author} {\bibfnamefont
  {M.}~\bibnamefont {He}}, \bibinfo {author} {\bibfnamefont {R.}~\bibnamefont
  {Lortz}}, \bibinfo {author} {\bibfnamefont {Y.}~\bibnamefont {Cai}}, \bibinfo
  {author} {\bibfnamefont {N.}~\bibnamefont {Wang}}, \bibinfo {author}
  {\bibfnamefont {T.}~\bibnamefont {Zhang}}, \bibinfo {author} {\bibfnamefont
  {H.}~\bibnamefont {Zhang}}, \bibinfo {author} {\bibfnamefont
  {Z.}~\bibnamefont {Tang}}, \bibinfo {author} {\bibfnamefont {P.}~\bibnamefont
  {Sheng}}, \bibinfo {author} {\bibfnamefont {H.}~\bibnamefont {Muramatsu}},
  \bibinfo {author} {\bibfnamefont {Y.~A.}\ \bibnamefont {Kim}}, \bibinfo
  {author} {\bibfnamefont {M.}~\bibnamefont {Endo}}, \bibinfo {author}
  {\bibfnamefont {P.~T.}\ \bibnamefont {Araujo}},\ and\ \bibinfo {author}
  {\bibfnamefont {M.~S.}\ \bibnamefont {Dresselhaus}},\ }\bibfield  {title}
  {\bibinfo {title} {{Superconductivity in Bundles of Double-Wall Carbon
  Nanotubes}},\ }\href {https://doi.org/10.1038/srep00625} {\bibfield
  {journal} {\bibinfo  {journal} {Scientific Reports}\ }\textbf {\bibinfo
  {volume} {2}},\ \bibinfo {pages} {625} (\bibinfo {year} {2012})}\BibitemShut
  {NoStop}%
\bibitem [{\citenamefont {Berdiyorov}\ \emph {et~al.}(2009)\citenamefont
  {Berdiyorov}, \citenamefont {Elmurodov}, \citenamefont {Peeters},\ and\
  \citenamefont {Vodolazov}}]{Berdiyorov2009}%
  \BibitemOpen
  \bibfield  {author} {\bibinfo {author} {\bibfnamefont {G.~R.}\ \bibnamefont
  {Berdiyorov}}, \bibinfo {author} {\bibfnamefont {A.~K.}\ \bibnamefont
  {Elmurodov}}, \bibinfo {author} {\bibfnamefont {F.~M.}\ \bibnamefont
  {Peeters}},\ and\ \bibinfo {author} {\bibfnamefont {D.~Y.}\ \bibnamefont
  {Vodolazov}},\ }\bibfield  {title} {\bibinfo {title} {{Finite-size effect on
  the resistive state in a mesoscopic type-II superconducting stripe}},\ }\href
  {https://doi.org/10.1103/PhysRevB.79.174506} {\bibfield  {journal} {\bibinfo
  {journal} {Phys. Rev. B}\ }\textbf {\bibinfo {volume} {79}},\ \bibinfo
  {pages} {174506} (\bibinfo {year} {2009})}\BibitemShut {NoStop}%
\bibitem [{\citenamefont {Berdiyorov}\ \emph {et~al.}(2014)\citenamefont
  {Berdiyorov}, \citenamefont {Harrabi}, \citenamefont {Oktasendra},
  \citenamefont {Gasmi}, \citenamefont {Mansour}, \citenamefont {Maneval},\
  and\ \citenamefont {Peeters}}]{Berdiyorov2014}%
  \BibitemOpen
  \bibfield  {author} {\bibinfo {author} {\bibfnamefont {G.}~\bibnamefont
  {Berdiyorov}}, \bibinfo {author} {\bibfnamefont {K.}~\bibnamefont {Harrabi}},
  \bibinfo {author} {\bibfnamefont {F.}~\bibnamefont {Oktasendra}}, \bibinfo
  {author} {\bibfnamefont {K.}~\bibnamefont {Gasmi}}, \bibinfo {author}
  {\bibfnamefont {A.~I.}\ \bibnamefont {Mansour}}, \bibinfo {author}
  {\bibfnamefont {J.~P.}\ \bibnamefont {Maneval}},\ and\ \bibinfo {author}
  {\bibfnamefont {F.~M.}\ \bibnamefont {Peeters}},\ }\bibfield  {title}
  {\bibinfo {title} {{Dynamics of current-driven phase-slip centers in
  superconducting strips}},\ }\href
  {https://doi.org/10.1103/PhysRevB.90.054506} {\bibfield  {journal} {\bibinfo
  {journal} {Phys. Rev. B}\ }\textbf {\bibinfo {volume} {90}},\ \bibinfo
  {pages} {054506} (\bibinfo {year} {2014})}\BibitemShut {NoStop}%
\bibitem [{\citenamefont {Xi}\ \emph {et~al.}(2016)\citenamefont {Xi},
  \citenamefont {Wang}, \citenamefont {Zhao}, \citenamefont {Park},
  \citenamefont {Law}, \citenamefont {Berger}, \citenamefont {Forr{\'o}},
  \citenamefont {Shan},\ and\ \citenamefont {Mak}}]{Xi2016}%
  \BibitemOpen
  \bibfield  {author} {\bibinfo {author} {\bibfnamefont {X.}~\bibnamefont
  {Xi}}, \bibinfo {author} {\bibfnamefont {Z.}~\bibnamefont {Wang}}, \bibinfo
  {author} {\bibfnamefont {W.}~\bibnamefont {Zhao}}, \bibinfo {author}
  {\bibfnamefont {J.-H.}\ \bibnamefont {Park}}, \bibinfo {author}
  {\bibfnamefont {K.~T.}\ \bibnamefont {Law}}, \bibinfo {author} {\bibfnamefont
  {H.}~\bibnamefont {Berger}}, \bibinfo {author} {\bibfnamefont
  {L.}~\bibnamefont {Forr{\'o}}}, \bibinfo {author} {\bibfnamefont
  {J.}~\bibnamefont {Shan}},\ and\ \bibinfo {author} {\bibfnamefont {K.~F.}\
  \bibnamefont {Mak}},\ }\bibfield  {title} {\bibinfo {title} {{Ising pairing
  in superconducting NbSe$_2$ atomic layers}},\ }\href
  {https://doi.org/10.1038/nphys3538} {\bibfield  {journal} {\bibinfo
  {journal} {Nature Physics}\ }\textbf {\bibinfo {volume} {12}},\ \bibinfo
  {pages} {139} (\bibinfo {year} {2016})}\BibitemShut {NoStop}%
\bibitem [{\citenamefont {de~la Barrera}\ \emph {et~al.}(2018)\citenamefont
  {de~la Barrera}, \citenamefont {Sinko}, \citenamefont {Gopalan},
  \citenamefont {Sivadas}, \citenamefont {Seyler}, \citenamefont {Watanabe},
  \citenamefont {Taniguchi}, \citenamefont {Tsen}, \citenamefont {Xu},
  \citenamefont {Xiao},\ and\ \citenamefont {Hunt}}]{delaBarrera2018}%
  \BibitemOpen
  \bibfield  {author} {\bibinfo {author} {\bibfnamefont {S.~C.}\ \bibnamefont
  {de~la Barrera}}, \bibinfo {author} {\bibfnamefont {M.~R.}\ \bibnamefont
  {Sinko}}, \bibinfo {author} {\bibfnamefont {D.~P.}\ \bibnamefont {Gopalan}},
  \bibinfo {author} {\bibfnamefont {N.}~\bibnamefont {Sivadas}}, \bibinfo
  {author} {\bibfnamefont {K.~L.}\ \bibnamefont {Seyler}}, \bibinfo {author}
  {\bibfnamefont {K.}~\bibnamefont {Watanabe}}, \bibinfo {author}
  {\bibfnamefont {T.}~\bibnamefont {Taniguchi}}, \bibinfo {author}
  {\bibfnamefont {A.~W.}\ \bibnamefont {Tsen}}, \bibinfo {author}
  {\bibfnamefont {X.}~\bibnamefont {Xu}}, \bibinfo {author} {\bibfnamefont
  {D.}~\bibnamefont {Xiao}},\ and\ \bibinfo {author} {\bibfnamefont {B.~M.}\
  \bibnamefont {Hunt}},\ }\bibfield  {title} {\bibinfo {title} {{Tuning Ising
  superconductivity with layer and spin-orbit coupling in two-dimensional
  transition-metal dichalcogenides}},\ }\href
  {https://doi.org/10.1038/s41467-018-03888-4} {\bibfield  {journal} {\bibinfo
  {journal} {Nature Communications}\ }\textbf {\bibinfo {volume} {9}},\
  \bibinfo {pages} {1427} (\bibinfo {year} {2018})}\BibitemShut {NoStop}%
\bibitem [{\citenamefont {Xing}\ \emph {et~al.}(2017)\citenamefont {Xing},
  \citenamefont {Zhao}, \citenamefont {Shan}, \citenamefont {Zheng},
  \citenamefont {Zhang}, \citenamefont {Fu}, \citenamefont {Liu}, \citenamefont
  {Tian}, \citenamefont {Xi}, \citenamefont {Liu}, \citenamefont {Feng},
  \citenamefont {Lin}, \citenamefont {Ji}, \citenamefont {Chen}, \citenamefont
  {Xue},\ and\ \citenamefont {Wang}}]{Xing2017}%
  \BibitemOpen
  \bibfield  {author} {\bibinfo {author} {\bibfnamefont {Y.}~\bibnamefont
  {Xing}}, \bibinfo {author} {\bibfnamefont {K.}~\bibnamefont {Zhao}}, \bibinfo
  {author} {\bibfnamefont {P.}~\bibnamefont {Shan}}, \bibinfo {author}
  {\bibfnamefont {F.}~\bibnamefont {Zheng}}, \bibinfo {author} {\bibfnamefont
  {Y.}~\bibnamefont {Zhang}}, \bibinfo {author} {\bibfnamefont
  {H.}~\bibnamefont {Fu}}, \bibinfo {author} {\bibfnamefont {Y.}~\bibnamefont
  {Liu}}, \bibinfo {author} {\bibfnamefont {M.}~\bibnamefont {Tian}}, \bibinfo
  {author} {\bibfnamefont {C.}~\bibnamefont {Xi}}, \bibinfo {author}
  {\bibfnamefont {H.}~\bibnamefont {Liu}}, \bibinfo {author} {\bibfnamefont
  {J.}~\bibnamefont {Feng}}, \bibinfo {author} {\bibfnamefont {X.}~\bibnamefont
  {Lin}}, \bibinfo {author} {\bibfnamefont {S.}~\bibnamefont {Ji}}, \bibinfo
  {author} {\bibfnamefont {X.}~\bibnamefont {Chen}}, \bibinfo {author}
  {\bibfnamefont {Q.-K.}\ \bibnamefont {Xue}},\ and\ \bibinfo {author}
  {\bibfnamefont {J.}~\bibnamefont {Wang}},\ }\bibfield  {title} {\bibinfo
  {title} {{Ising Superconductivity and Quantum Phase Transition in Macro-Size
  Monolayer NbSe$_2$}},\ }\href {https://doi.org/10.1021/acs.nanolett.7b03026}
  {\bibfield  {journal} {\bibinfo  {journal} {Nano Letters}\ }\textbf {\bibinfo
  {volume} {17}},\ \bibinfo {pages} {6802} (\bibinfo {year}
  {2017})}\BibitemShut {NoStop}%
\bibitem [{\citenamefont {Ambegaokar}\ and\ \citenamefont
  {Baratoff}(1963{\natexlab{a}})}]{AB}%
  \BibitemOpen
  \bibfield  {author} {\bibinfo {author} {\bibfnamefont {V.}~\bibnamefont
  {Ambegaokar}}\ and\ \bibinfo {author} {\bibfnamefont {A.}~\bibnamefont
  {Baratoff}},\ }\bibfield  {title} {\bibinfo {title} {{Tunneling Between
  Superconductors}},\ }\href {https://doi.org/10.1103/PhysRevLett.10.486}
  {\bibfield  {journal} {\bibinfo  {journal} {Phys. Rev. Lett.}\ }\textbf
  {\bibinfo {volume} {10}},\ \bibinfo {pages} {486} (\bibinfo {year}
  {1963}{\natexlab{a}})}\BibitemShut {NoStop}%
\bibitem [{\citenamefont {Ambegaokar}\ and\ \citenamefont
  {Baratoff}(1963{\natexlab{b}})}]{ABerratum}%
  \BibitemOpen
  \bibfield  {author} {\bibinfo {author} {\bibfnamefont {V.}~\bibnamefont
  {Ambegaokar}}\ and\ \bibinfo {author} {\bibfnamefont {A.}~\bibnamefont
  {Baratoff}},\ }\bibfield  {title} {\bibinfo {title} {{Tunneling Between
  Superconductors (Erratum)}},\ }\href
  {https://doi.org/10.1103/PhysRevLett.11.104} {\bibfield  {journal} {\bibinfo
  {journal} {Phys. Rev. Lett.}\ }\textbf {\bibinfo {volume} {11}},\ \bibinfo
  {pages} {104} (\bibinfo {year} {1963}{\natexlab{b}})}\BibitemShut {NoStop}%
\end{thebibliography}%


\clearpage
\newpage

\onecolumngrid
\begin{center}
\textbf{\large Supplementary Material: Supercurrent and phase slips in a ballistic carbon nanotube embedded into a van der Waals heterostructure}
\end{center}

\twocolumngrid
\beginsupplement

 \section{Carbon nanotube growth from thin Co films using chemical vapor deposition}
An array of reference markers (20~nm Re$_{70}$Mo$_{30}$, 5~nm Pt, 50~$\mu$m spacing) is patterned on a dedicated p-doped Si chip with a 285~nm thick capping layer of thermal SiO$_2$ using standard electron beam lithography (EBL), sputtering and electron beam evaporation techniques. 
These markers serve as reference for subsequent fabrication of catalyst dots, localization of grown carbon nanotubes (CNTs) by scanning electron microscopy (SEM), as well as for the CNT pick-up process.\par 
Next, catalyst dot structures (diameter 4~$\mu$m, nominal thickness 1~nm) are prepared from thin Co films by EBL and thermal evaporation.
CNTs are grown~\cite{Blien2018} by chemical vapor depositions (CVD) at 950~$^\circ$C for 20~minutes. Methane is used as carbon feed stock (flow 10~sccm) and hydrogen as carrier gas (flow 20~sccm). The interior of the reaction chamber is kept oxygen-free during heating and cooling by an argon flow of 1500~sccm.\par 
CNTs are located on the substrate with respect to the marker array by SEM imaging (aperture 30~$\mu$m, acceleration voltage 2~kV, working distance 6~mm, in-lens detector). Under these conditions, typically a 10 up to several 100~$\mu$m long CNT grows from about every hundredth catalyst point, often relatively straight and in the gas flow direction.
\section{Details about the stamping process}
\subsection{Exfoliation on PDMS}
The fabrication of a van der Waals stack starts with a few mm large bulk crystal. 
Commercially available blue Nitto-tape (SPV 224P, as of now simply 'blue tape') is used for efficient thinning of the bulk crystal down to the few-layer regime. 
By bringing the blue tape and the bulk crystal into contact and retracting again, we can pull off several macroscopic portions of the bulk crystal onto the blue tape. This produces the so-called 0$^\text{th}$ crystal generation. In order to further thin down the crystals we repeatedly (3-5 times) cleave the crystals using fresh blue tape every time. This results in the subsequent 1$^\text{st}$ to 5$^\text{th}$ generations. 
The blue tape with most promising generation is put onto ant then retracted quickly from a rectangular ($\sim8$~mm$\times8$~mm) thin film of polydimethylsiloxane (PDMS). 
The PDMS was prepared on a glass slide before.
%
Typically, we observe the best results with the 2$^\text{nd}$ or 3$^\text{rd}$ generations. 
Lastly, we search under the optical microscope for flakes suiting our purposes with respect to size, thickness and other surrounding flakes. 
%
The NbSe$_2$ crystals were bought from \textit{hq$^+$graphene}.
\subsection{Direct 2D material stamping onto Si/SiO\textsubscript{2} substrate}
We follow the approach of Ref.~\cite{Castellanos_Gomez_2014}.
As target substrate for our stamping process we use $4.5$~mm$\times 4.5$~mm chips cut from  p-Si/SiO$_2$ wafer. The oxide layer is 285~nm thick.
On these chips a rectangular array of alignment markers with a $x,y$-spacing of 50~$\mu$m is patterned by means of standard EBL and thermal metal evaporation (1~nm Cr, 30~nm Au).
For stamping, we fix the wafer chips on common double-sided tape on top of a $x,y,z$-piezo positioner. 
To visually monitor the process, we use an adjustable zoom lens paired with a camera livestreaming to a TV screen~\cite{Castellanos_Gomez_2014}.
The glass slide with the desired flake on the PDMS is mounted into a micro-manipulator stage and positioned just above the target substrate. The actual stamping is controlled by micrometer screws, while the piezo-stack is used for rough adjustments. 
To begin the stamping process, we approach the substrate with the glass slide/PDMS, stopping it when contact is established. The glass slide is then further lowered, in a slow fashion in order to keep control of the movement of the contact area front, which moves across the substrate.
This front line is moved across the desired flake. Once the whole flake has made good contact with the substrate the motion of the contact area front is reversed by retracting the glass slide. 
When the whole flake sticks,  the PDMS film is retracted completely. One can use this technique to subsequently stamp multiple flakes on top of each other.
In our devices we use this direct stamping method to prepare the graphite/bottom hBN layers for the CNT.
\begin{figure*}[tb]
\centering
\includegraphics[width=\textwidth]{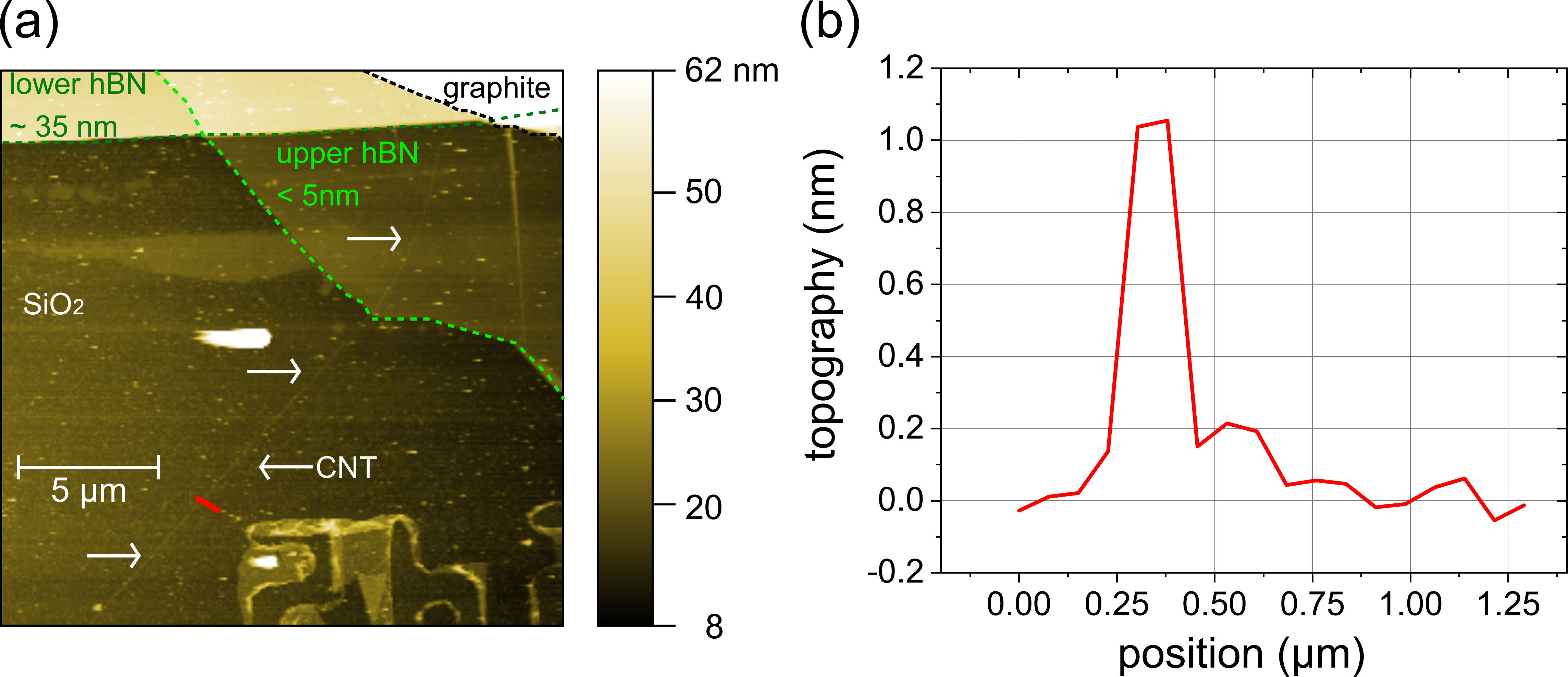}
\caption{ (a) Atomic force microscopy (AFM) topography scan showing the carbon nanotube (CNT, white arrows) transferred  onto the graphite and the lower hBN flake. The AFM scan has been performed just after the stamping process and before the fabrication of the electrodes. The scan is performed at the periphery of the device, where the (CNT) is not covered by the van der Waals crystals and is thus accessible to AFM topography. The boundary of the crystals (whose thickness is reported) is indicated with a dashed line. (b) Line cut of the AFM topography data corresponding to the red segment in (a). We deduce a CNT diameter of approximately $1.2\pm 0.5$~nm. Despite the sizable error bar, we can exclude a thick bundle of many CNTs.}
\label{fig:AFM-images}
\end{figure*}
\subsection{Pick-up technique for 2D crystals and CNTs}
We follow the approach of Ref.~\cite{Wang614}.
The pick-up technique can be used to transfer a sequentially picked-up 2D flakes from a Si/SiO$_2$ substrate to another. In our approach we apply this method also to CNTs.
A lens-shaped PDMS stamp is covered by a thin polycarbonate (PC) film. The PDMS lens is fabricated by placing single drops of PDMS base material and PDMS curing agent onto a glass slide and baking out for 10~minutes on a hot plate set to 150~$^\circ$C. 
We fabricate the PC film by sandwiching a handful of drops of 5~\%~PC/chloroform solution (by weight) between two glass slides. Once all the air between the slides is expelled, we smoothly and quickly slide of the top glass slide and a thin and even film of PC/chloroform solution remains on the bottom slide. 
To drive off the solvent we bake out the slide on a hotplate set to 150~$^\circ$C for 5~minutes. The resulting PC film can be stored in air until use. 
We use common scotch tape to transfer the PC film onto the PDMS lens. First, we punch a hole in the scotch tape with an office perforator. Second, using the scotch tape we lift the PC film from the glass slide so, that the film spans over the hole without tearing. We fix the scotch tape with the PC film onto the PDMS lens using more tape.

To provide temperature control during the procedure we used a home-made chip holder made of copper on top of the piezo stack. It  includes a DC powered heating resistor and a thermometer. An orifice connected to a vacuum pump provides the necessary suction to fix the substrates to the copper block.
We load the glass slide with the PC layer facing down into the micromanipulator stage and position the tip of the PDMS lens above the flake/CNT we want to pick up. 
Before starting the pick-up process, the temperature of the sample holder on the piezo stack must be below 50~$^\circ$C as otherwise the PC layer suffers from premature sagging.
First, we approach in such a way, that the tip of the PC layer/PDMS lens touches the substrate next to the flake we want to pick up (about 100~$\mu$m away). 
The desired flake/CNT shall not be touched with the PC layer at this stage. 
Then the copper block is heated up up to 124~$^\circ$C. 
The temperature increase will soften the PC and it will bend downwards. Thus, the contact area will increase into a circular shape and the meniscus will flow completely over the flake/CNT. 
The glass slide/PC layer is retracted in a smooth motion and with it the flake/CNT as soon as the $124\pm 2~^\circ$C are reached.
If desired, other flakes can be picked-up with this very stack by repeating the steps mentioned above. 	
The drop-off is essentially a complete melting of the PC layer. First, flake(s)/CNT are aligned on the PC layer with the target substrate or graphite/bottom hBN bed. 
Second, the flake(s)/CNT are brought in contact with the substrate. Then the copper block is heated up to $\sim 180~^\circ$C to melt the PC layer. When the PC layer is fully molten, one can retract the glass slide and the van der Waals stack remains under a circular PC coating. This PC coating can afterwards be dissolved in chloroform. Afterwards, an intermediate AFM imaging determines the exact position of transferred CNT.
Since CNTs are invisible to the eye, it is absolutely necessary when transferring CNTs to mark down their exact position of the CNT relative to landmarks in the PC layer/PDMS lens and referencing the SEM pictures of the CNT. We achieve a consistent lateral precision of $~5~\mu$m using our setup. This tolerance has to be kept in mind, when selecting suitable flakes.	
\section{Edge contact fabrication}

For the fabrication of edge contacts on CNTs, NbSe$_2$ and graphite/graphene we use a common etching process based on reactive ion etching. This process is routinely used to make contacts on high-mobility hBN -encapsulated graphene devices. An application of this method to CNTs (where the edge is limited to a handful of atoms) is reported by Huang \textit{et al.} in Ref.~\cite{HuangCNTedge}

The geometry of the electrodes is defined by EBL. To access the hBN-encapsulated targed material we need to etch several nanometers (typically 10-20~nm) of hBN first.
The etching process is realized in an Oxford Plasmlab 80 system. 
Before the actual etching process, we condition the chamber with an oxygen plasma (gas flow 100~sccm, chamber  pressure 100~mTorr, power 200~W, process time 10~min). For the actual etching we use a mixture of CHF$_3$/O$_2$ gas with flow rates of 40~sccm/6~sccm (chamber pressure 35~mTorr, power 35~W, etching rate $~0.45$~nm/s). 
Immediately thereafter, the etched sample is mounted in a UHV chamber (pressure $1\times10^{-6}$~mbar). A 1~nm high chromium layer is applied as an adhesion promoter, and 100~nm gold is applied as a contact metal. 
Before performing the lift-off, place the sample in acetone for one hour.

\section{Atomic force microscopy images}
Figure~\ref{fig:AFM-images} shows two false color AFM images of our sample. 
From Fig.~\ref{fig:AFM-images}(a) we can determine the NbSe$_2$ flake (purple bordered) thickness to $\lesssim 3$~nm and from Fig.~\ref{fig:AFM-images}(b) the thickness of the lower hBN flake to $\sim 35$~nm (dark green bordered), upper hBN flake to $\sim 5$~nm (light green bordered) and the graphite flake $\sim 40$~nm (gray bordered).
The CNT (white appearing line) is also visible. It is marked with black arrows. 
%
%
%
From the available AFM scans, we can estimate a CNT diameter of $0.7\pm0.5$~nm. Despite the large error bar, we can conclude that (i) the diameter is well within the typical diameter distribution (1-2~nm) for our CVD growth process; (ii) though the presence of few bundled CNTs cannot be ruled out, the presence of many CNTs can definitely be excluded. 

The current-independent value of the conductance between contacts 2-3  at low bias is roughly 120~$e^2/h$, which corresponds to a resistance of   215~$\Omega$. Since at least 80~$\Omega$ are due to the cable resistance, we are left with more than 190~$e^2/h$ ($\equiv 135$~$ \Omega$) of sample conductance, which we attribute to the Au/NbSe$_2$ edge contact (typically of the order of hundreds of ohms for several micrometer-long edge contacts).

\section{Measurement setup}
Transport measurements were performed in a dilution refrigerator with a base temperature of 30~mK.
We performed two-terminal measurements using a combined voltage excitation of a Stanford Lock-in amplifier SR830 (AC $20\mu V$, 27.77~Hz) mixed with a DC Signal from a Yokogawa 7651  DC voltage source (voltage division 1000/100).
The current is converted back into a voltage signal and amplified by a Femto Ampere Current Amplifier DDPCA-300. AC component is detected with the same lock-in amplifier, whereas the DC part is measured with a Keysight 3458A Digital Multimeter. The graphite gate is biased with another Yokogawa 7651.
%
%
%
%
\section{Voltage vs current bias dependence of the resistive peaks}
\begin{figure*}[tb]
\includegraphics[width=\textwidth]{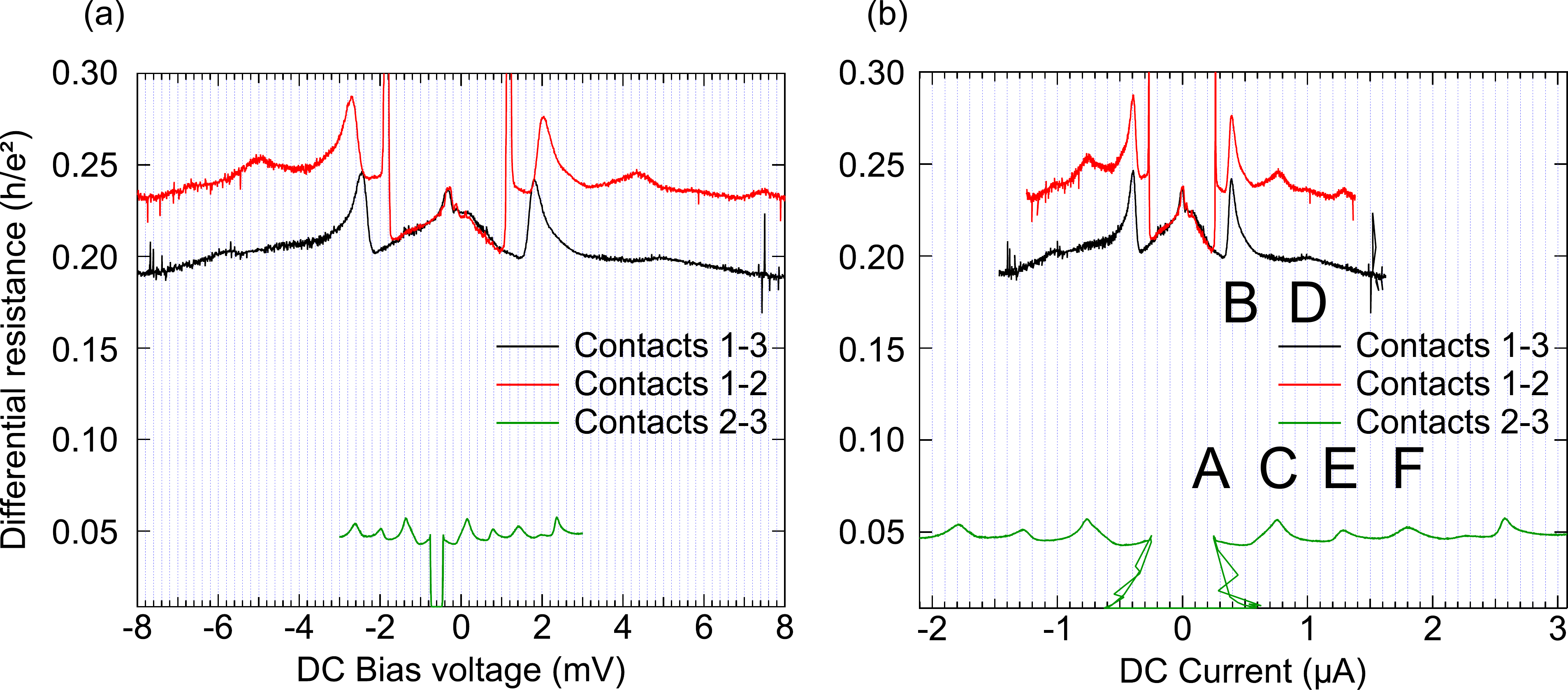}
\caption{Differential resistance as a function of (a) voltage bias and (b) current bias for contact configuration 1-3 (red), 1-2 (black) and 2-3 (green). Measurements were performed at 30~mK in absence of magnetic field. Various features of the different traces are aligned on top of each other only when plotted as a function of current bias.}
\label{fig:current-voltage-bias}
\end{figure*}
Figure~\ref{fig:current-voltage-bias} shows that the differential resistance for contacts 1-2 (red), 1-3 (black) and 2-3 (green) as a function of (a) voltage bias and (b) current bias. Resistance peaks in the different traces are aligned on top of each other only when the resistance is plotted versus current and not voltage bias. This clearly indicates that the resistive features are triggered by current.
\begin{figure*}[tb]
\includegraphics[width=\textwidth]{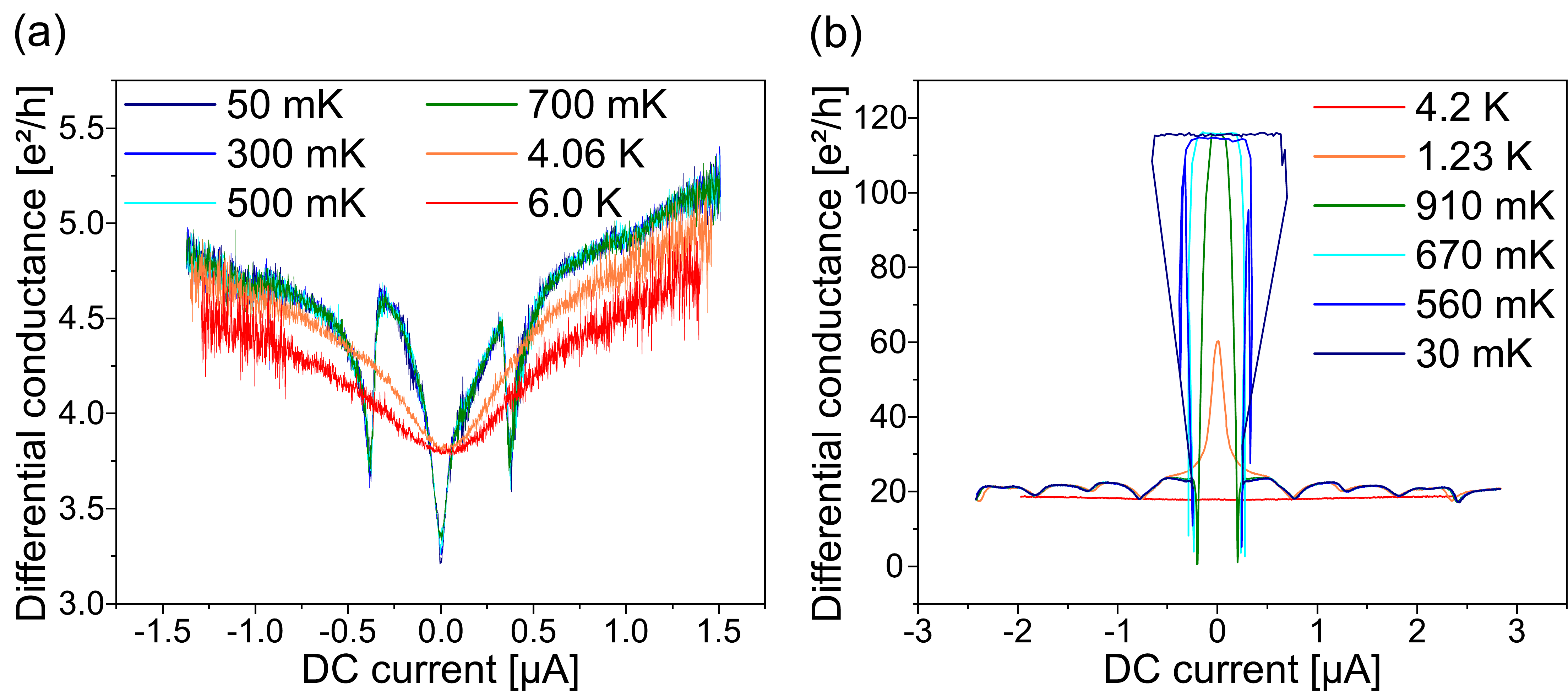}
\caption{Temperature dependence of the differential resistance for contacts (a) 1-3 and (b) 2-3, respectively. Measurements were taken in absence of magnetic field.}
\label{fig:T-dependence}
\end{figure*}

\section{Temperature dependence of $G_{1-3}(I)$ and $G_{2-3}(I)$}
Figure~\ref{fig:T-dependence} shows the temperature dependence of the differential conductance $G(I)$ for the contact configurations 1-3 and 2-3, respectively. These measurements were taken in absence of magnetic field.
Figure~\ref{fig:T-dependence}(a) shows that the minima in the differential conductance $G$ are independent of the DC current up to at least 700~mK and disappear at latest 4.06~K. The temperature range between 800~mK and 4.2~K is difficult to access in a dilution refrigerator and therefore only these few measurement data are available for this measurement.
Figure~\ref{fig:T-dependence}(b) shows that the high differential conductance of $\sim 120$~$e^2/h$ for small DC currents has a pronounced temperature dependence. The maximum conductance is maintained up to a temperature of 910~mK, even if the width of this maximum becomes smaller and at 560~mK two side maxima are visible. In contrast, the minima for $I_\text{DC}>0.6$~$\mu$A are independent for low temperatures up to more than 1~K. At 4.2~K these also disappear.

\end{document}